\documentclass[preprint,12pt,authoryear]{elsarticle}

\usepackage[linkcolor=black,colorlinks=true,urlcolor=blue]{hyperref}
\usepackage{mathtools}

\usepackage{color,xcolor,ucs}
\usepackage{mathtools}   \usepackage{tikz} 
\usepackage{ amssymb }
\usepackage{extarrows} 
\usepackage{pgf,tikz}
\usepackage{float}
\usetikzlibrary{positioning}
\usetikzlibrary{shapes.geometric}
\usetikzlibrary{shapes.misc}
\usetikzlibrary{arrows}
\usepackage{caption}
\usepackage{mathrsfs}
\usetikzlibrary{arrows,shapes,automata,backgrounds,petri,positioning}
\usetikzlibrary{decorations.pathmorphing}
\usetikzlibrary{decorations.shapes}
\usetikzlibrary{decorations.text}
\usetikzlibrary{decorations.fractals}
\usetikzlibrary{decorations.footprints}
\usetikzlibrary{shadows}
\usetikzlibrary{calc}
\usetikzlibrary{spy}
\usepackage{amsmath}
\usepackage{array}
\usepackage{ amssymb }
\usepackage{braket}
\usepackage{qcircuit}
\usepackage{soul}
\usepackage{braket} 
\usepackage{relsize}

\usepackage{amsmath}
\usepackage{ amssymb }
\usepackage{braket}
\usepackage{qcircuit}
\usepackage{soul}
\usepackage{braket}

\begin{document}

\begin{frontmatter}



\title{Quantum inverse scattering for the 20-vertex model up to Dynkin automorphism: 3D Poisson structure, triangular height functions, weak integrability} 


\author{Pete Rigas}

\affiliation{ 
         Newport Beach, CA 92625 pbr43@cornell.edu
        }

\begin{abstract}
      We initiate a novel application of the quantum inverse scattering method for the 20-vertex model, building upon seminal work from Faddeev and Takhtajan on the study of Hamiltonian systems. In comparison to a previous work of the author in late $2023$ which characterized integrability of a Hamiltonian flow for the 6-vertex model from integrability of inhomogeneous limit shapes, formalized in a work of Keating, Reshetikhin and Sridhar, notions similar to those of  integrability can be realized for the 20-vertex model by studying new classes of higher-dimensional L-operators. Such L-operators provided by Boos and colleagues have algebraic, combinatorial, and geometric, qualities, all of which impact leading order approximations of correlations, products of L-operators, the transfer matrix, and the quantum monodromy matrix. \footnote{\textbf{MSC Class}: 81U40; 34L25; 60K35; 82B20; 82B27} 
\end{abstract}

\begin{graphicalabstract}
\end{graphicalabstract}




\end{frontmatter}



\section{Introduction}

\subsection{Overview}

The six-vertex (6V) model, originally introduced to study thermodynamic properties of ice {[34]}, has received significant attention besides seminal computations for the residual free energy on the torus {[30]}, ranging from quantification of crossing probabilities through a Russo-Seymour-Welsh (RSW) formalism {[10]}, computation of the free energy from the conjectured phase diagram into ferroelectric, and antiferroelectric, phases {[11]}, integrability and spin-boson models {[1},{16},{31},{41]}, exact solvability {[4},{5},{6},{9]}, staggering {[17},{23]}, limit shapes {[35]}, spin-spin, and ordinary, correlations {[8},{26},{27]}, and various extensions of Russo-Seymour-Welsh results to the six-vertex model under sloped boundary conditions, in addition to obtaining crossing probability estimates for the the Ashkin-Teller and generalized random-cluster models over the strip {[36]}. To continue exploring several aspects of interest, which can share connections with behaviors observed in other models of statistical mechanics, from the influence of boundary conditions {[13},{42},{46]}, discrete holomorphicity from the Cauchy-Riemann equations {[40]}, symmetry {[18},{29},{32]}, to the Bethe ansatz {[5},{12]}, we pursue new applications of the quantum-inverse scattering approach, {[16]}, which has lead to many applications for studying lattice models that are directly solvable and integrable.  In a previous application of the quantum inverse scattering approach for the inhomogeneoous six-vertex model, {[41]}, computations for establishing that the inhomogeneous Hamiltonian flow, from results stating that the inhomogeneous limit shapes are integrable, is also integrable, include: identifying a suitable L-operator, obtaining a system of relations between the Poisson bracket and entries of the transfer matrices (see the system of relations provided in \textit{2.4} of {[41]}, as an adaptation of the system provided in seminal work on Hamiltonian methods in {[16]}), applying properties of the Poisson bracket to simplify expressions obtained from the system of relations parametrized in entries of the transfer matrix, along with discussing more general features of the Poisson bracket of entries of transfer and monodromy matrices together - the Poisson structure. 

Albeit the fact that integrability has been rigorously established for the 6-vertex model from the notion of Poisson commutativity, [24], knowledge of integrability for all vertex models is still not complete, and determining whether related versions of integrability that have previously been developed for uncolored vertex models can also be applied to colored vertex models is of interest. More generally, determining whether a lattice model is completely solvable, and integrable, is of value for determining whether the correlations of the model can be precisely quantified, in addition to the fact that suitable action-angle variables exist for the system, which can be expressed by making use of standard properties of the Poisson bracket. Despite the fact that the Poisson bracket appears in any argument for establishing that a desired integrability property holds (from adaptations of the method presented in {[16]}), additional complications from such arguments arise from: identifying suitable action-angle coordinates; determining whether the candidate action-angle coordinates in question can be approximated from properties of the Poisson bracket; establishing relations between entries of the quantum monodromy and transfer matrices; quantifying the influence of boundary conditions of the vertex model for correlations in the finite volume bulk. 

\noindent For the purposes of the quantum inverse scattering method, the presence of boundary conditions is significant in that the boundary conditions determine the structure of the L-operators; the L-operator is used from the very beginning of arguments using the quantum inverse scattering approach for relations that are obtained from entries of the quantum monodromy and transfer matrices. For colored vertex models, to further develop notions of integrability and exact solvability, a variant of the quantum inverse scattering approach is possible, consisting of: modifications to the R matrix for constructing colored vertex models; studying the dependence between the structure of the L-operators and sloped boundary conditions that can be imposed on the six-vertex model; further examinations of connections between Representation Theory, Algebraic Combinatorics, and Algebraic Geometry, through the formulation of the action of the Dynkin automorphism on diagrams, [5], examined in this work.

To contribute across areas of research interest within the field of Integrable Probability, we define several objects relating to those manipulated for the quantum inverse scattering approach for the uncolored six-vertex model with homogeneities {[41]}. Besides other structures which have previously been studied for vertex models and other settings of Statistical Mechanics {{[37}},{{38}},{{39}},{{40}},{{43]}}, we raise the attention of the reader to possible applications of the forthcoming computations to other models of interest. Despite the fact that the $D^{(2)}_2$ spin chain and uncolored, inhomogeneous six-vertex model alike were not studied with the root density approach as the spin chain was, defining a Hamiltonian flow for the colored, inhomogeneous six-vertex model, as has been done from a conjectured raised in {[24]} regarding  the very interesting question as to whether integrability of the Hamiltonian flow can be inferred from the result on integrability of inhomogeneous limit shapes, can be applied towards understanding additional conditions on integrability of a Hamiltonian flow for the inhomogeneous six-vertex model. In other models, while the notion of integrability can be utilized to describe the model's behavior in the presence of disorder, other models of interest in Statistical Mechanics can be analyzed with parafermionic observables for bosons and fermions alike {[40]}, scaling limits {[43]}, a quadrichotomy of four possible behaviors for crossing probabilities {[35]}, tiling models {[48]}, finite temperature properties {[43]}, Painleve equations {[45]}, sharpness of the phase transition for the two-dimensional Gaussian free field {[38]}, Schur functions {[33]}, rationality of vertex models {[1]}, along with several other connections {[41}, {42]}, pertaining to the encoding of boundary conditions, and to eigenvalue attraction.

\subsection{This paper's contributions}

This paper develops a novel construction for the transfer and quantum monodromy matrices of the 20-vertex model through a factorization of the Universal R-matrix. In comparison to the construction of other R-matrices in the Mathematical Physics literature, the Universal R-matrix simultaneously incorporates interactions from Classical and Quantum Physics through a parameterization of several algebraic and combinatorial factors. Albeit the fact that the effect of the boundary conditions can be approximated through the asymptotic expansion of products of $L$ operators, being able to relate how more complicated interactions arise in three-dimensions, in comparison to those over two-dimensions, is of interest to formulate. The action-angle coordinates for the 20-vertex model, as a higher dimensional analog of those for the 6-vertex model, are dependent upon a generalization of the Quantum interactions captured through domain-wall L operators for the 6-vertex model. We compare the notion of integrability from action-angle coordinates in the 6-vertex and 20-vertex models, through the notion of weak integrability described in .

\subsection{Paper Organization}

To investigate which notions of integrability from the two-dimensional six-vertex model transfer, if at all, to the three-dimensional six-vertex model, we obtain approximations of the weak infinite volume limit of the three-dimensional transfer matrix. In comparison to the presence of inhomogeneities that appear in the spectral parameters of the two-dimensional transfer matrix which imply that an integrability property of the Hamiltonian flow holds from integrability conditions imposed on the limit shape, the following approach consists of: identifying recursive relations when one spectral parameter is varied while another spectral parameter is held constant; extrapolating such a set of relations for varying another spectral parameter while two spectral parameters are held constant; identifying a broader set of relations from which computations from the Poisson bracket can be executed from different pairs of entries of the three-dimensional transfer matrix. In two-dimensions, the quantum inverse scattering type approach, as an adaptation of seminal work appearing in {[16]} for the nonlinear Schrodinger's equation with Hamiltonian methods. Within this framework, before inhomogeneities are imposed in the six-vertex model, the absence of the nonlinear Schrodinger's equation implies that some portions of the argument must be developed for the solutions to the Bethe equations obtained from the algebraic Bethe ansatz. In three-dimensions, the construction of the universal R-matrix, along with the universal R-matrix itself, no longer satisfy the Bethe equations which raises differences in the same portion of the arguments as for those of the inhomogeneous six-vertex model {[41]}.

The forthcoming approach obtains a broader set of relations than those previously described in computations with the Poisson bracket of entries of the transfer matrix from that obtained in {[41]}. The larger number of terms which appear in one set of relations expressed in terms of the Poisson bracket implies that properties that it satisfies, anticommutativity, bilinearity, Leibniz' rule, and the Jacobi identity, continue to play a large role in approximations of each term within the set of relations. Moreover, as a higher dimensional analog, the set of relations with the Poisson bracket can be expressed in terms of the set of relations obtained from one Poisson bracket up to permutation of entries of the transfer matrix. For the 20-vertex model there exists a set of relations that one can map into from the set of $81$ relations stated in \textit{1.5.1}. In the next section, when performing the lowest order approximations of the three-dimensional transfer matrix for a few terms at a time, we denote the product of $\mathcal{I}^i$, where $\mathcal{I}^i$ denotes the product of $i$, and even an arbitrary number, of L-operators that are used to first approximate the entries of the transfer matrix, for every $- N \leq i \leq 0$. After having obtained a set of relations for the entries of the three-dimensional transfer matrix, the 
three-dimensional quantum monodromy matrix,

{\small \begin{align*}
 T^{3D}_{a,b} {\tiny \big(    \big\{ u_i \big\} , \big\{ v^{\prime}_j  \big\} , \big\{ w^{\prime\prime}_k \big\}     \big)}  :   \textbf{C}^3 \otimes \big( \textbf{C}^3 \big)^{\otimes ( |N| + ||M||_1 )}  \longrightarrow   \textbf{C}^3 \otimes \big( \textbf{C}^3 \big)^{\otimes ( |N| + ||M||_1 )}  \\  \mapsto \overset{\underline{M}}{\underset{k=0}{\prod}} \text{ } \overset{0}{\underset{j=-N}{\prod}}    \bigg\{   \mathrm{diag} {\tiny \big( } \mathrm{exp} {\tiny \big(  \alpha \big(i, j,k \big)  \big)}   , \mathrm{exp} {\tiny \big(  \alpha \big( i,j,k\big)  \big) }  , \mathrm{exp} {\tiny \big(  \alpha \big( i, j,k \big)  \big) \big) }  R_{ia,jb,kc} {\tiny \big( u - u_i ,  u^{\prime} - v^{\prime}_j , w-w^{\prime\prime}_k \big) } \bigg\}   \text{, } 
\end{align*} }

\noindent is further analyzed from the set of $81$ relations with the Poisson bracket, which parallels the definition of the two-dimensional quantum monodromy matrix,

{\small \begin{align*}
 T_a {\tiny \big( u , \big\{ v_k \big\} , H , 0 \big) } : \textbf{C}^2 \otimes \big( \textbf{C}^2 \big)^{\otimes |N|} \longrightarrow \textbf{C}^2 \otimes { \big( \textbf{C}^2 \big)^{\otimes |N|} }      \mapsto    \overset{1}{\underset{i=-N}{\prod}}   \bigg\{  \mathrm{diag} {\tiny \big( \mathrm{exp} {\tiny \big( 2H \big)}  ,  \mathrm{exp} {\tiny \big( 2 H \big)}   \big) }     \\ \times   R_{ia} {\tiny \big( u - v_i \big)   }    \bigg\}    \text{, } 
\end{align*} }

\noindent defined with factors of the two-dimensional R-matrix,

{\small \[
R \equiv R {\tiny \big( u , H , V \big)}  \equiv 
  \begin{bmatrix}
      a \text{ }  \mathrm{exp} {\tiny \big(  H + V \big) }    & 0 & 0 & 0  \\
    0 & b \text{ } \mathrm{exp} {\tiny \big( H - V \big)}  & c & 0  \\0 & c & b \text{ }  \mathrm{exp} {\tiny \big( - H + V \big)}  & 0 \\ 0 & 0 & 0 & a \text{ }  \mathrm{exp} {\tiny \big( - H - V \big) }   \text{ } 
  \end{bmatrix} \text{. } 
\] } 

\subsection{Objects of the six-vertex and inhomogeneous six-vertex models}

\noindent We provide an overview of quantities for defining the Hamiltonian flow of the inhomogeneous model, including, modifications to the R and L-operators from those of the 6-vertex model. Recall that some six-vertex configuration $\omega$ on $\textbf{T}_N = \big(V {\tiny \big( } \textbf{T}_N {\tiny \big)} , E {\tiny \big(} \textbf{T}_N {\tiny \big)}  \big)$, for $N$ even, is weighed according to the following product over each possible vertex type,

{\small \begin{align*}
  w_{6V} {\tiny \big(\omega \big)}  \equiv w {\tiny \big(\omega \big)} =        a_1^{n_1} a_2^{n_2} b_1^{n_3} b_2^{n_4} c_1^{n_5} c_2^{n_6} \underset{ \{ c_1 \equiv c_2 \equiv c \} }{\underset{\{ b_1 \equiv b_2 \equiv b\} }{\underset{\{ a_1 \equiv a_2 \equiv a\} }{\overset{\mathrm{Isotropic}}{\Longleftrightarrow}} }}     a^{n_1+n_2}  b^{n_3+n_4}  c^{n_5+n_6}    \text{, }   
\end{align*} }

\noindent for $a_1,a_2,b_1,b_2,c_1,c_2 \geq 0$, and each $n_i$ indicates the number of vertices of type $i$, from which the corresponding probability measure takes the form,

{\small \begin{align*}
  \textbf{P}_{\textbf{T}_N} {\tiny \big[      \omega         \big]}   \equiv \textbf{P}{\tiny \big[   \omega     \big]}     =  \frac{w_{6V}{\tiny \big(\omega \big)}}{Z_{\textbf{T}_N}}   \equiv \frac{w{\tiny \big(\omega \big)}}{Z_{\textbf{T}_N}}   \text{, }  \\
\end{align*} }

\noindent where the $Z_{\textbf{T}_N} \equiv Z = \sum_{\omega \in \Omega} w {\tiny \big(\omega \big)}$ is the partition function (see \textit{Figure 1} for the arrow representation of the six-vertex model, and \textit{Figure 2} for the line representation of the six-vertex model).  The disorder parameter for the six-vertex model takes the form,

{\small \begin{align*}
    \Delta \equiv \frac{a_1 a_2 + a_3 a_4 - a_5 a_6}{2 \sqrt{a_1 a_2 a_3 a_4}}     \text{. } 
\end{align*} }

\noindent Under other assumptions on the choice of weights rather than the isotropic choice above,

{\small \begin{align*}
  w_{\mathrm{6V}} {\tiny \big( \omega \big)}  \equiv w {\tiny \big( \omega \big)}  \equiv a_1^{n_1} a_2^{n_2} b_1^{n_3} b_2^{n_4} c_1^{n_5} c_2^{n_6}  \text{, }
\end{align*}}

\noindent in the presence of inhomogeneities the six-vertex model, and stochastic six-vertex model, have limit shapes satisfying several properties, from satisfying the Euler Lagrange equations to having a Hamiltonian structure, with the interactions that can be expressed as the maximum over $\pm$, with, {[24]},

{\small \begin{align*}
  \mathcal{H}_u {\tiny \big( q , H \big)}  \equiv \underset{\pm}{\mathrm{max}}    \text{ } \mathcal{H}^{\pm}_u {\tiny \big( q , H \big)}  \equiv \underset{\pm}{\mathrm{max}}  \big\{     \pm H + l_{\pm} + \int_C  \psi^{\pm}_u {\tiny \big( \alpha \big)}  \rho {\tiny \big( \alpha \big) }   \text{ } \mathrm{d} \alpha   \big\}   \text{, } 
\end{align*} }

\noindent where $l_{-} \equiv \mathrm{log} \big\{  \mathrm{sinh} {\tiny \big( \eta - u \big)} \big\}$, $l_{+} \equiv \mathrm{log} \big\{ \mathrm{sinh} \text{ }  u\big\}$, and the density,

{\small \begin{align*}
  \rho {\tiny \big( \alpha \big)}  \equiv       \# \big\{    \text{Bethe roots along contours } C        \big\}   \equiv \underset{\alpha > 0 }{\bigcup}  \big\{ \alpha : \alpha \cap C \neq \emptyset \big\} \equiv  \#  \big\{ \alpha : \alpha \cap C \neq \emptyset \big\}     \text{, } 
\end{align*} }

\noindent for the Bethe equations,

{\small \begin{align*}
     \overset{N}{\underset{k=1}{\prod}}  \big\{   {\mathrm{sinh} {\tiny \big(   \frac{\eta}{2} + i \alpha_j - v_k  \big)} } \big/ {\mathrm{sinh} {\tiny \big(     \frac{\eta}{2} - i \alpha_j + v_k    \big) } }  \big\}  \\        = \mathrm{exp} {\tiny \big( 2 H N          \big) } \cdot \overset{n}{\underset{m=1, m \neq j}{\prod} }        \big\{      {\mathrm{sinh} {\tiny \big(   i \big( \alpha_j - \alpha_m \big)+ \eta    \big)}} \big/ {\mathrm{sinh} {\tiny \big(    i \big( \alpha_j - \alpha_m \big) - \eta      \big)} }  \big\}        \text{, } 
\end{align*} } 

\noindent which has solutions ${\tiny \big\{ \alpha_i \big\}}$, for,

{\small \begin{align*}
     \psi_{+} {\tiny \big( \alpha + i u\big) =    \mathrm{log}\big[  \frac{\mathrm{sinh} \big(    \frac{\eta}{2} + u - i \alpha    \big) }{\mathrm{sinh} \big(    \frac{\eta}{2} - u + i \alpha    \big) }  \big]  }   \text{, }  \\    \psi_{-} {\tiny \big( \alpha + i u\big)     =  \mathrm{log} \big[  \frac{\mathrm{sinh} \big(     \frac{3 \eta}{2} - u + i \alpha     \big) }{\mathrm{sinh} \big(    u - \frac{\eta}{2} - i \alpha     \big) }     \big]  }  \text{. } 
\end{align*} }

\noindent Explicitly, under $\pm$ the Hamiltonian flow takes the form, {[24]},

{\small \[
\mathcal{H}^{\pm}_u {\tiny \big( q , H \big)}  \equiv  \text{ } 
\left\{\!\begin{array}{ll@{}>{{}}l}       + \Rightarrow    \underset{+}{\mathrm{max}}  \big\{     H + \mathrm{log} \text{ } \mathrm{sinh} \text{ }  u + \int_C  \mathrm{log}\big[  \frac{\mathrm{sinh} (    \frac{\eta}{2} + u - i \alpha  - iu  ) }{\mathrm{sinh} (    \frac{\eta}{2} - u + i \alpha - iu    ) }  \big]    \rho {\tiny \big( \alpha \big) }   \text{ } \mathrm{d} \alpha   \big\}    
 \text{, } \\  
 - \Rightarrow   \underset{-}{\mathrm{max}}  \big\{  -   H + \mathrm{log} \text{ } \mathrm{sinh} \big( \eta - u \big) + \int_C  \mathrm{log} \big[  \frac{\mathrm{sinh} (     \frac{3 \eta}{2} - u + i \alpha - iu     ) }{\mathrm{sinh} (    u - \frac{\eta}{2} - i \alpha  - iu    ) }     \big]      \rho {\tiny \big( \alpha \big)  }  \text{ } \mathrm{d} \alpha   \big\}      \text{, }  \\
\end{array}\right.
\] }

\noindent Integrability of the Hamiltonian flow for the inhomogeneous six-vertex model relates to analyses of the transfer matrix,

{\small \begin{align*}
   t {\tiny \big( u , \big\{ v_k \big\} , H , V \big) }  : \big( \textbf{C}^2 \big)^{\otimes N} \longrightarrow \big( \textbf{C}^2 \big)^{\otimes N}   \text{, }  
\end{align*} }

\noindent quantum monodromy matrix,

{\small \begin{align*}
 T_a {\tiny \big( u , \big\{ v_k \big\} , H , 0 \big) } : \textbf{C}^2 \otimes \big( \textbf{C}^2 \big)^{\otimes N} \longrightarrow \textbf{C}^2 \otimes \big( \textbf{C}^2 \big)^{\otimes N}      \mapsto    \overset{N}{\underset{i=1}{\prod}}   \big\{  \mathrm{diag} {\tiny \big( \mathrm{exp} \big( 2H \big) ,  \mathrm{exp} \big( 2 H \big)  \big) }       R_{ia} {\tiny \big( u - v_i \big) }    \big\}     \text{, } 
\end{align*} }

\noindent Notions from Representation Theory emerge within the field of Integrable Probability through assumptions that the R matrix satisfies before solutions, amongst several other properties, of the Bethe and Yang Baxter equations can be discussed. The R-matrix above for the inhomogeneous six-vertex model satisfies the Yang-Baxter equation,

{\small \begin{align*}
R_{12} {\tiny \big( u \big) } R_{13} {\tiny \big( u + v \big)} R_{23} {\tiny \big( v \big) } =  R_{23} {\tiny \big( v \big) } R_{13} {\tiny \big( u + v \big)}  R_{12} {\tiny \big( u \big) }  \text{. } 
\end{align*} } 

\noindent for $0 < u < \eta$, and $0 < v < \eta$, give suitable $\eta > 0$. Furthermore, under a paramterization of the weights for the six-vertex model in the presence of two external fields, $H$ and $V$,

{\small \begin{align*}
   a_1 \equiv      a \text{ }  \mathrm{exp} {\tiny \big(  H + V \big)}   \text{, } \\ a_2 \equiv   a  \text{ }  \mathrm{exp} {\tiny \big( - H - V \big) }  \text{, } \\  b_1 \equiv  \text{ }  \mathrm{exp} {\tiny \big( H - V \big) }  \text{, } \\ b_2 \equiv \text{ }  \mathrm{exp} {\tiny \big( - H + V \big) }  \text{, } \\ c_1 \equiv  c \lambda  \text{, } \\ c_2 \equiv c \lambda^{-1} \text{, } 
\end{align*} }

\noindent for $a_1 \equiv a_2 \equiv a$, $b_1 \equiv b_2 \equiv b$, $c_1 \equiv c_2 \equiv c$, and $\lambda \geq 1$. To study other characteristics of the six-vertex model, ranging from connections between uncolored and colored vertex models to the influence of boundary conditions, we specify objects for a quantum-inverse scattering method (often times denoted QISM)-type argument, under the presence of inhomogeneities in the uncolored six-vertex model, one obtains expressions for Poisson brackets between entries of the transfer matrix. In the uncolored setting, the L-operator for which such computations are carried out takes the form,

{\small \[
       L_{\alpha , k   } {\tiny \big( \lambda_{\alpha} , v_{k} \big)   }  \equiv 
  \begin{bmatrix}
     \mathrm{sin} {\tiny \big( \lambda_{\alpha} - v_k + \eta \sigma^z_k \big)     }   &    \mathrm{sin} {\tiny \big( 2 \eta \big) } \sigma^{-}_k    \\
      \mathrm{sin} {\tiny \big( 2 \eta \big) } \sigma^{+}_k     &   \mathrm{sin} {\tiny \big( \lambda_{\alpha} - v_k - \eta \sigma^z_k \big)     } 
  \end{bmatrix}  \text{, } 
\] }

\noindent as defined in {[7]}, which was suggested to be a direction of interest in {[24]}, where, corresponding to the $k$ th horizontal line, $\alpha$ th vertical line, with $L_{\alpha,k} \curvearrowright \big(  \text{vertical space } \otimes  \text{horizontal space } \big) $. In order to obtain an asymptotic expansion for the entries of the transfer and quantum monodromy matrices from L-operators expressed in three Pauli bases, we introduce quantities $D^i_j$ and $D^i_{j+1}$ into entries of the L-operator in three dimensions.

\subsection{Triangular ice of the 20-vertex model}

We formally define the 20-vertex model under domain-wall boundary conditions, which have the below weights ${\tiny w_0, \cdots, w_6}$. If the ambient state space of configurations is taken to be over $\textbf{T}$ instead of over $\textbf{Z}^2$, the 6-vertex model can be defined in a very similar way to its lower dimensional counterpart. Instead of six possible configurations, over $\textbf{T}$ there are twenty possible configurations (such as in the line representation of configurations for the 6-vertex model, the weights, and accompanying sample space, of the 20-vertex model), in addition to the fact that, before enforcing an isotropic parameter choice on the six-vertex weights $a_1, a_2, b_1, b_2, c_1, c_2$ with $a \equiv a_1 \equiv a_2$, $b \equiv b_1 \equiv b_2$ and $c \equiv c_1 \equiv c_2$, the 20-vertex weights take the form, {[15]},

{\small  \begin{align*}
     w_0 \equiv   a_1 a_2 a_3   \text{, } 
\\ \\ 
    w_1 \equiv   b_1 a_2 b_3   \text{, } \\ \\  w_2    \equiv   b_1 a_2 c_3    \text{, } \\ \\ w_3 \equiv       a_1 b_2 b_3 + c_1 c_2 c_3  \text{, } \\ \\  w_4 \equiv    c_1 a_2 a_3        \text{, } \\ \\              w_5 \equiv   b_1 c_2 a_3   \text{, } \end{align*}
  
  \begin{align*} w_6 \equiv  b_1 b_2 a_3    \text{. } 
\end{align*} }

\noindent for the 20-vertex probability measure, from the $6$ possible weights above, $\textbf{P}^{20V}_{\textbf{T}}{\tiny \big[      \omega         \big] }    \equiv \textbf{P}^{20V} {\tiny \big[   \omega     \big] }      =  {w_{20V} {\tiny \big(\omega \big)} \big/  }{Z^{20V}_{\textbf{T}}} \equiv {w {\tiny \big(\omega \big)}} \big/ {Z_{\textbf{T}}} $, for some ${\tiny \omega \in \Omega^{20V}}$.

 \bigskip 
\noindent Besides the representation of the state space of the three-dimensional six-vertex model reproduced from {[15]} another figure is reproduced from {[15]} which demonstrates how pairs of configurations, two at a time, are assigned each one of the possible weights $w_i$ defined above. As alluded to in {[15]}, there exists many special cases of the six parameters for the twenty-vertex model, which differ other special parameterisations of weights that are dependent upon whether the disorder parameter $\Delta$ satisfies $\Delta = -1$, $\Delta< -1$, or $ \Delta -1< $ and $\Delta <1$, as is the case for the 6-vertex model {[11]}.

  \bigskip 

\noindent In addition to the square and triangular ice models, one must also introduce the Poisson bracket. As an operation that takes two arguments, the Poisson bracket $\big\{ \cdot , \cdot \big\}$ satisfies the following set of properties, given test functions $f$, $g$ and $h$,

\begin{itemize}
    \item [$\bullet$] \textit{Anticommutativity}. $\big\{ f, g \big\}  =  - \big\{ g , f \big\} $

    \item[$\bullet$] \textit{Bilinearity}. For real $a,b$, $\big\{ af + bg , h \big\} = a \big\{ f ,h \big\} + b \big\{ g , h \big\},$ and $\big\{ h , af + bg \big\} = a \big\{ h , f \big\} + b \big\{ h , g \big\} $

    \item[$\bullet$] \textit{Leibniz' rule}. $\big\{ fg , h \big\} = \big\{ f , h \big\} g + f \big\{ g , h \big\}$

    \item[$\bullet$] \textit{Jacobi identity}. $\big\{ f , \big\{ g , h \big\} \big\} + \big\{ g , \big\{ h , f \big\} \big\} + \big\{ h , \big\{ f , g \big\} \big\} = 0$ \end{itemize}

\subsection{Representation theoretic assumptions of the universal R-matrix}

\noindent Considering L-operators of the form above is significant for: identifying ways in which generalizations of vertex models depend upon the dimension of the ambient lattice in which interactions occur; computing

\noindent sets of functional relations that that are parameterized in entries of the L-operator; formulating connections from underlying representations of the R matrix, with combinatorial and geometric consequences. As the ambient space of the six-vertex model becomes three-dimensional, additional interactions appear in the L-operator and universal R-matrix, {[5]},

{\small \[
R = R_{\leq \delta} R_{\sim \delta} R_{\geq \delta} K \text{, }
\]}

\noindent where,

{\small \begin{align*}
\left\{\!\begin{array}{ll@{}>{{}}l}    R_{\leq \delta} \equiv   \underset{m \in \textbf{N}}{\underset{\gamma \in \Delta_{+} ( A)}{\prod}}       \mathrm{exp}_q    \big[   {\tiny \big( q - q^{-1} \big)     }        s^{-1}_{m , \gamma} e_{\gamma + m \delta} \otimes f_{\gamma + m \delta}          \big] \text{, } \\ \\  R_{\sim \delta}  \equiv     \mathrm{exp} \big[  {\tiny \big( q - q^{-1} \big) } \underset{m \in \textbf{Z}^{+}}{\sum}  \text{ }   \overset{r}{\underset{i , j =1}{\sum}}      u_{m,ij} e_{m \delta , \alpha_i } \otimes f_{m \delta , \alpha_j }        \big]    \text{, }  \\ \\ 
 R_{\geq \delta} \equiv           \underset{m \in \textbf{N}}{\underset{\gamma \in \Delta_{+} ( A)}{\prod}}       \mathrm{exp}_q    \big[   {\tiny \big( q - q^{-1} \big)    }         s^{-1}_{m , \delta - \gamma} e_{\delta - \gamma + m \delta} \otimes f_{ \delta - \gamma + m \delta}          \big]      \text{, } \\ \\  K \equiv   \mathrm{exp} \big[   \hbar     \overset{r}{ \underset{i , j =1 }{\sum} } {\tiny \big(  b_{ij} h_{\alpha_i} \otimes h_{\alpha_j} \big)}    \big]    \text{, } \end{array}\right. 
\end{align*} }

\noindent which satisfies a set of equations different from those provided in the statement of the Bethe equations whose solutions are analyzed from the Bethe ansatz, for the q-exponential,

{\small \begin{align*}
   \mathrm{exp}_q {\tiny \big[ z \big] } \equiv  \overset{+ \infty}{\underset{n=0}{\sum}}      \frac{z^n}{{\tiny \big[ n \big]_q}  !}      \equiv      \overset{+ \infty}{\underset{n=0}{\sum}}   z^n  \bigg[ \frac{{\tiny \big( 1 - q \big)^n} }{\underset{1 \leq i \leq n}{\prod} {\tiny \big( 1 - q^i \big)    }   }   \bigg]              \text{. }
\end{align*} } 

\noindent and for objects satisfying the conditions,

{\small \begin{align*}
   \left\{\!\begin{array}{ll@{}>{{}}l}     \frac{q- q^{-1}}{q^{h_{\gamma + m \delta}} - q^{-h_{\gamma + m \delta}}}     \big[ e_{\gamma +  m \delta} ,  f_{\gamma + m \delta} \big]   =   s_{m , \gamma }        \text{, } \\         \\                t_{m , ij}   =  {\tiny \big( -1 \big)^{m ( 1 - \delta_{ij} )}  }      m^{-1} \big[ m a_{ij} \big]_q           \Longleftrightarrow             \frac{1}{e_{\alpha_i + ( m + n ) \delta}}   \big[   e_{\alpha_i + m \delta} , e_{n \delta , \alpha_j}  \big]_q                 = t_{n , ij}     \text{, } \\   \\    \frac{q- q^{-1}}{q^{h_{\delta - \gamma + m \delta}} - q^{-h_{\delta - \gamma + m \delta}}}     \big[ e_{\delta - \gamma +  m \delta} ,  f_{ \delta - \gamma + m \delta} \big]   =   s_{m , \delta -  \gamma }            \text{, } \\ \\   B \equiv  \big\{    b_{ij}  \big\}_{i,j \in \textbf{N}} \Longleftrightarrow     \exists \text{ }  B : B A \equiv \textbf{I}           \text{. } \end{array}\right.  
\end{align*} }

\noindent The quantities $\gamma$ are taken over a well-defined root system.

\bigskip

\noindent Three-dimesional L-operators for the 20-vertex model exhibit additional interaction terms appearing from the span of the third row and column. From the previously identified parametrization for weights for the six-vertex model in the presence of two external fields, [24], computations with the Poisson bracket result from a set of sixteen relations,

{\tiny \[
\left\{\!\begin{array}{ll@{}>{{}}l} \boxed{(1)}:      \big\{  A \big( u \big)        , A \big( u^{\prime} \big)   \big\} 
\text{ , } \\  \boxed{(2)}:  \big\{         A \big( u \big)        ,        B \big( u^{\prime} \big)     \big\}  \text{ , } \\  \boxed{(3)}:   \big\{   A \big( u \big)       ,  C \big( u^{\prime} \big) \big\} 
 \text{, }   \\  \boxed{(4)}: \big\{   A \big( u \big)       ,  D \big( u^{\prime} \big) \big\}   \text{ , } \\ \boxed{(5)} : \big\{ B \big( u \big) , A \big( u^{\prime} \big) \big\} \text{, } \\ \boxed{(6)}: \big\{ B \big( u \big) , B \big( u^{\prime} \big) \big\} \text{, } \\ \boxed{(7)}: \big\{ B \big( u \big) , C \big( u^{\prime} \big) \big\} \text{, } \\ \boxed{(8)}: \big\{ B \big( u \big)  , D \big( u^{\prime} \big)   \big\} \text{, } 
 \\  \boxed{(9)}:  \big\{ C \big( u \big)  , A \big( u^{\prime} \big)   \big\} \text{, } \\ \boxed{(10)}: \big\{ C \big( u \big) , B \big( u^{\prime} \big) \big\} \text{, } \\ \boxed{(11)}: \big\{ C \big( u \big) , C \big( u^{\prime} \big) \big\} \text{, } \\ \boxed{(12)}: \big\{ C \big( u \big) , D \big( u^{\prime} \big) \big\} \text{, } \\  \boxed{(13)}: \big\{ D \big( u \big)  , A \big( u^{\prime} \big)  \big\} \text{, } \\ \boxed{(14)}: \big\{  D \big( u \big)  , B \big( u^{\prime} \big)  \big\} \text{, } \\ \boxed{(15)}: \big\{  D \big( u \big)  , C \big( u^{\prime} \big) \big\}  \text{, }  \\ \boxed{(16)}: \big\{  D \big( u \big)  ,       D \big( u^{\prime} \big)  \big\} \text{, }
\end{array}\right.
\] }

\noindent from the tensor product, which in the coordinates $u,u^{\prime}$ is equivalent to,

{\small \[
\big\{ T_a {\tiny \big( u , \big\{ v_k \big\} \big) } \overset{\bigotimes }{,}   T_a {\tiny \big( u^{\prime} , \big\{ v^{\prime}_k \big\} \big) }   \big\} = \bigg\{    \begin{bmatrix} 
A {\tiny \big( u \big) } & B {\tiny \big( u \big) } \\ C {\tiny \big( u \big)} & {\tiny D \big( u \big)  } 
\end{bmatrix}\overset{\bigotimes }{,}   \begin{bmatrix} 
A {\tiny \big( u^{\prime} \big) } & B {\tiny \big( u^{\prime} \big)}  \\ C {\tiny \big( u^{\prime} \big)}  & D {\tiny \big( u^{\prime} \big)   } 
\end{bmatrix}  \bigg\}  \text{, } 
\] }

\noindent which by the definition of Poisson bracket of the tensor product of the two reduced monodromy matrices, equals,

{\small \[
   \bigg\{    \bigg[  r_{a,+}        {\tiny \big( v_k - v^{\prime}_k \big) } \begin{bmatrix} 
A {\tiny \big( u \big)}  & B {\tiny \big( u \big) } \\ C {\tiny \big( u \big)}  & D {\tiny \big( u \big) }   
\end{bmatrix}  \bigg]   \otimes    \begin{bmatrix} 
A {\tiny \big( u^{\prime} \big)  } & B {\tiny \big( u^{\prime} \big) }  \\ C {\tiny \big( u^{\prime} \big)} & D {\tiny \big( u^{\prime} \big)}   
\end{bmatrix}  \bigg\} - \bigg\{  \begin{bmatrix} 
A {\tiny \big( u \big)}  & B {\tiny \big( u \big) } \\ C {\tiny \big( u \big)}  & D {\tiny \big( u \big) }   
\end{bmatrix} \otimes   \begin{bmatrix}
A {\tiny \big( u^{\prime} \big)}  & B {\tiny \big( u^{\prime} \big) } \\ C {\tiny \big( u^{\prime} \big)}  & D {\tiny \big( u^{\prime} \big) }   
\end{bmatrix}  \] \[ \times  r_{a,-}      {\tiny \big( v_k - v^{\prime}_k  \big)  }             \bigg] \bigg\} \text{. } 
\] }

\noindent where the quantities in the tensor products above, besides the approximation for entries of the two-dimensional transfer matrix take the form,

{\small \begin{align*}
     r_{a,+}      {\tiny \big( v_k - v^{\prime}_k  \big) }  =       \underset{y \longrightarrow + \infty}{\mathrm{lim}}  \big\{ E^{\mathrm{6V}} {\tiny \big(   u^{\prime},  v^{\prime}_k - v_k  \big)}  \otimes  \big[                E^{\mathrm{6V}} {\tiny \big(   u^{\prime},  v^{\prime}_k - v_k   \big)  }     r_a {\tiny \big(  v_k - v^{\prime}_k  \big)   }         \big]  \big\}  \text{, } \\  \\      r_{a,-}      {\tiny \big(  v_k - v^{\prime}_k  \big) }  =      \underset{y \longrightarrow - \infty}{\mathrm{lim}} \big\{ E^{\mathrm{6V}} {\tiny \big(   u^{\prime}  ,  v^{\prime}_k - v_k \big)   }    \otimes  \big[                E^{\mathrm{6V}} {\tiny \big(  u^{\prime}  ,  v^{\prime}_k - v_k \big)  }    r_a {\tiny \big(  v_k - v^{\prime}_k  \big) }       \big] \big\}    \text{, } 
\end{align*}}

\noindent for,

{\small \begin{align*}
     E^{\mathrm{6V}} {\tiny \big( v  - v_k  , x \big) } = \mathrm{exp} \big[             \mathrm{coth} {\tiny \big( \frac{\eta}{2} + i \alpha_j - v_k \big)    }         \big]   \text{. }
\end{align*} } 

\noindent For the following objects, denote $\underline{u} \neq \underline{u^{\prime}}$ as spectral parameters over $\textbf{T}$. Given a transfer matrix,
{\small \[
\begin{bmatrix}
 A {\tiny \big( \underline{u} \big)}  & D {\tiny \big( \underline{u} \big)} & G {\tiny \big( \underline{u} \big)} \\ B {\tiny \big( \underline{u} \big)} & E {\tiny \big( \underline{u} \big)} & H {\tiny \big( \underline{u} \big)}  \\ C {\tiny \big( \underline{u} \big)}   &  F {\tiny \big( \underline{u} \big)} & I {\tiny \big( \underline{u} \big) } 
\end{bmatrix}  \text{, } 
\]} 

\noindent one obtains a larger set of relations, $81$ for the three-dimensional case in comparison to $16$ from the two-dimensional case. Over three dimensions, the transfer matrix from the factorization of the Universal R-Matrix implies that the Poisson brackets take the form,

{\tiny  \[
\left\{\!\begin{array}{ll@{}>{{}}l} \boxed{(1)}:      \big\{  A \big( \underline{u} \big)        , A \big( \underline{u^{\prime}} \big)   \big\} 
\text{, } &  \boxed{(22)}: \big\{ C \big( \underline{u} \big) , D \big( \underline{u^{\prime}} \big) \big\} \text{, }  &  \boxed{(43)}: \big\{ E \big( \underline{u} \big) , G \big( \underline{u^{\prime}} \big) \big\} \text{, }  \\  \boxed{(2)}:  \big\{         A \big( \underline{u} \big)        ,        B \big( \underline{u^{\prime}} \big)     \big\}  \text{
, } &  \boxed{(23)}:   \big\{  C \big( \underline{u} \big) , E \big( \underline{u^{\prime}} \big)  \big\}  \text{, } &  \boxed{(44)}: \big\{ E \big( \underline{u} \big) , H \big( \underline{u^{\prime}} \big) \big\} \text{, }   \\  \boxed{(3)}:   \big\{   A \big( \underline{u} \big)       ,  C \big( \underline{u^{\prime}} \big) \big\} 
 \text{, } &  \boxed{(24)}: \big\{ C \big( \underline{u} \big) , F \big( \underline{u^{\prime}} \big) \big\} \text{, } & \boxed{(45)}: \big\{ E \big( \underline{u} \big) , I \big( \underline{u^{\prime}} \big) \big\} \text{, } \\ \boxed{ (4)}: \big\{   A \big( \underline{u} \big)       ,  D \big( \underline{u^{\prime}} \big) \big\}   \text{, } & \boxed{(25)}: \big\{ C \big( \underline{u} \big) , G \big( \underline{u^{\prime} } \big) \big\} \text{, } &  \boxed{(46)}: \big\{ F \big( \underline{u} \big) ,     A \big( \underline{u^{\prime }} \big) \big\} \text{, } \\ \boxed{(5)}: \big\{ A \big( \underline{u} \big) , E \big( \underline{u^{\prime}} \big) \big\}  \text{,  } &   \boxed{(26)}: \big\{ C \big( \underline{u} \big) , H \big( \underline{u^{\prime}} \big) \big\} \text{, } & \boxed{(47)}:  \big\{ F \big( \underline{u} \big) , B \big( \underline{u^{\prime}}  \big) \big\} \text{, }  \\ \boxed{(6)} : \big\{ A \big( \underline{u} \big) , F \big( \underline{u^{\prime} } \big) \big\}  \text{ , }   & \boxed{(27)}: \big\{ C \big( \underline{u} \big) , I \big( \underline{u^{\prime}}  \big) \big\} \text{, } &     \boxed{(48)}: \big\{ F \big( \underline{u} \big) , C \big( \underline{u^{\prime}} \big) \big\} \text{, }     \\ \boxed{(7)}: \big\{ A \big( \underline{u} \big) , G \big( \underline{u^{\prime}} \big) \big\} \text{, }  &    \boxed{(28)}: \big\{ D \big( \underline{u} \big)  , A \big(\underline{u^{\prime}} \big)  \big\} \text{, } & \boxed{(49)}:   \big\{ F \big( \underline{u} \big) , D \big( \underline{u^{\prime}} \big) \big\} \text{, }  \\ \boxed{(8)}: \big\{ A \big( \underline{u} \big) , H \big( \underline{u^{\prime}} \big) \big\} &  \boxed{(29)}: \big\{  D \big( \underline{u} \big)  , B \big( \underline{u^{\prime}} \big)  \big\} \text{, } &   \boxed{(50)}: \big\{ F \big( \underline{u} \big) , E \big( \underline{u^{\prime}} \big) \big\} \text{, }  \\ \boxed{(9)}: \big\{ A \big( \underline{u} \big) , I \big( \underline{u^{\prime}} \big) \big\}  \text{, } &    \boxed{(30)}: \big\{  D \big( \underline{u} \big)  , C \big( \underline{u^{\prime}} \big) \big\} \text{, } &  \boxed{(51)}:   \big\{ F \big( \underline{u} \big) , F \big( \underline{u^{\prime}} \big) \big\} \text{, }   \\ \boxed{(10)} : \big\{ B \big( \underline{u} \big) , A \big( \underline{u^{\prime}} \big) \big\} \text{, } &    \boxed{(31)}: \big\{  D \big( \underline{u} \big)  ,       D \big( \underline{u^{\prime}} \big)  \big\} \text{, } &  \boxed{(52)}: \big\{ F \big( \underline{u} \big) , G \big( \underline{u^{\prime}} \big) \big\} \text{, }     \\ \boxed{(11)}: \big\{ B \big( \underline{u} \big) , B \big( \underline{u^{\prime}} \big) \big\} \text{, } &   \boxed{(32)} : \big\{ D \big( \underline{u} \big) , E \big( \underline{u^{\prime}}\big) \big\} \text{, } &   \boxed{(53)}: \big\{ F \big( \underline{u} \big) , H \big( \underline{u^{\prime}} \big) \big\} \text{, } \\  \boxed{(12)}: \big\{ B \big( \underline{u} \big) , C \big( \underline{u^{\prime}} \big) \big\} \text{, }  &  \boxed{(33)}: \big\{ D \big( \underline{u} \big) , F \big( \underline{u^{\prime}} \big) \big\} \text{, } &   \boxed{(54)}:  \big\{ F \big( \underline{u} \big) , I \big( \underline{u^{\prime}} \big) \big\} \text{, }  \\ \boxed{(13)}: \big\{ B \big( \underline{u} \big)  , D \big( \underline{u^{\prime}} \big)   \big\} \text{, } & \boxed{(34)}: \big\{ D \big( \underline{u} \big) , G \big( \underline{u^{\prime}} \big) \big\} \text{, }  &  \boxed{(55)}: \big\{ G \big( \underline{u} \big) , A \big( \underline{u^{\prime}} \big) \big\} \text{, }   \\ \boxed{(14)}: \big\{ B \big( \underline{u} \big) , E \big( \underline{u^{\prime}} \big) \big\}  \text{, } &   \boxed{(35)}: \big\{ D \big( \underline{u} \big) , H \big( \underline{u^{\prime}} \big) \big\} \text{, } &   \boxed{(56)}: \big\{ G \big( \underline{u} \big) , B \big( \underline{u^{\prime}} \big) \big\} \text{, } \\ \boxed{(15)}: \big\{ B \big( \underline{u} \big) , F \big( \underline{u^{\prime}} \big) \big\}  \text{, } &   \boxed{(36)}: \big\{ D \big( \underline{u} \big) , I \big( \underline{u^{\prime}} \big) \big\} \text{, } &  \boxed{(57)}:  \big\{ G \big( \underline{u} \big) , C \big( \underline{u^{\prime}} \big) \big\} \text{, } \\ \boxed{(16)}: \big\{ B \big( \underline{u} \big) , G \big( \underline{u^{\prime}} \big) \big\} \text{, } & \boxed{(37)}: \big\{ E \big( \underline{u} \big) , A \big( \underline{u^{\prime}} \big) \big\}  \text{, } &  \boxed{(58)}: \big\{ G \big( \underline{u} \big) , D \big( \underline{u^{\prime} }\big) \big\} \text{, } \\ \boxed{(17)}: \big\{ B \big( \underline{u} \big) , H \big( \underline{u^{\prime}} \big) \big\} \text{, } & \boxed{(38)}: \big\{ E \big( \underline{u} \big) , B \big( \underline{u^{\prime}} \big) \big\} \text{, } & \boxed{(59)} : \big\{ G \big( \underline{u} \big) , E \big( \underline{u^{\prime}} \big) \big\} \text{, } \\  \boxed{(18)}: \big\{ B \big( \underline{u} \big) , I \big( \underline{u^{\prime}} \big) \big\}   \text{, }  &    \boxed{(39)}: \big\{ E \big( \underline{u} \big) , C \big( \underline{u^{\prime}} \big) \big\} \text{, } & \boxed{(60)}:  \big\{ G \big( \underline{u} \big) , F \big( \underline{u^{\prime}} \big) \big\} \text{, }  \\  \boxed{(19)}:  \big\{ C \big( \underline{u} \big)  , A \big( \underline{u^{\prime}} \big)   \big\} \text{, } &    \boxed{(40)}:   \big\{ E \big( \underline{u} \big) , D \big( \underline{u^{\prime}} \big) \big\} \text{, } & \boxed{(61)}: \big\{ G \big( \underline{u} \big) , G \big( \underline{u^{\prime}} \big) \big\} \text{, } \\ \boxed{(20)}: \big\{ C \big( \underline{u} \big) , B \big( \underline{u^{\prime}} \big) \big\} \text{, } & \boxed{(41)}: \big\{ E \big( \underline{u} \big) , E \big( \underline{u^{\prime}} \big) \big\} \text{, } & \boxed{(62)}: \big\{ G \big( \underline{u} \big) , H \big( \underline{u^{\prime}} \big) \big\} \text{, } \\ \boxed{(21)}: \big\{ C \big( \underline{u} \big) , C \big( \underline{u^{\prime}} \big) \big\} \text{, }  &  \boxed{(42)}: \big\{ E \big( \underline{u} \big) , F \big( \underline{u^{\prime}} \big) \big\} \text{, } &  \boxed{(63)}: \big\{ G \big( \underline{u} \big) , I \big( \underline{u^{\prime}} \big) \big\} \text{, }
\\  \boxed{(64)}: \big\{ H \big( \underline{u} \big) , A \big( \underline{u^{\prime}} \big) \big\} \text{, } & \boxed{(70)}: \big\{ H \big(\underline{u} \big) , G \big( \underline{u^{\prime}}  \big) \big\} \text{, } & \boxed{(76)}: \big\{ I \big( \underline{u} \big) , D \big( \underline{u^{\prime}} \big) \big\} \text{, }  \\ \boxed{(65)}: \big\{ H \big( \underline{u} \big) , B \big( \underline{u^{\prime}} \big) \big\} \text{, } & \boxed{(71)}:  \big\{ H \big( \underline{u} \big) , H \big( \underline{u^{\prime} }\big) \big\} \text{, }      &   \boxed{(77)}: \big\{ I \big( \underline{u} \big) , E \big( \underline{u^{\prime}} \big) \big\} \text{, }   \\ \boxed{(66)}: \big\{ H \big( \underline{u} \big) , C \big( \underline{u^{\prime}} \big) \big\} \text{, } &  \boxed{(72)}: \big\{ H \big( \underline{u} \big) , I \big( \underline{u^{\prime}} \big) \big\} \text{, }         &       \boxed{(78)}: \big\{ I \big( \underline{u} \big) , F \big( \underline{u^{\prime}} \big) \big\} \text{, }    \\ \boxed{(67)} : \big\{ H \big( \underline{u} \big) , D \big( \underline{u^{\prime}} \big) \big\} \text{ , }  &  \boxed{(73)}: \big\{ I \big( \underline{u} \big) , A \big( \underline{u^{\prime}} \big) \big\} \text{, }        &   \boxed{(79)}: \big\{ I \big( \underline{u} \big) , G \big( \underline{u^{\prime}} \big) \big\} \text{, }  \\ \boxed{(68)}: \big\{ H \big( \underline{u} \big) , E \big( \underline{u^{\prime}} \big) \big\} \text{, }   &    \boxed{(74)}: \big\{ I \big( \underline{u} \big) , B \big( \underline{u^{\prime}} \big) \big\} \text{, }        &         \boxed{(80)}:  \big\{ I \big( \underline{u} \big) , H \big( \underline{u^{\prime}} \big) \big\} \text{, }          \\ \boxed{(69)}: \big\{ H \big( \underline{u} \big) , F \big( \underline{u^{\prime}} \big) \big\} \text{, }  &  \boxed{(75)}: \big\{ I \big( \underline{u} \big) , C \big( \underline{u^{\prime}} \big) \big\} \text{, }     & \boxed{(81)}: \big\{ I \big( \underline{u} \big) , I \big( \underline{u^{\prime}} \big) \big\} \text{, }   
\end{array}\right.
\] }



\noindent from the tensor product, which in the coordinates $u,u^{\prime}$ is equivalent to,

{\small \[
\big\{ \textbf{T} {\tiny \big( \underline{u}  , \big\{ u_i \big\} ,  \big\{ v_j \big\} , \big\{ w_k \big\}  \big)}  \overset{\bigotimes }{,}  \textbf{T} {\tiny \big( \underline{u^{\prime}} , \big\{ u^{\prime}_i \big\} ,  \big\{ v^{\prime}_j \big\} , \big\{ w^{\prime}_k \big\}   \big) }   \big\} = \bigg\{    \begin{bmatrix}
 A {\tiny \big( \underline{u} 
 \big)}  & D {\tiny \big( \underline{u} \big)}  & G {\tiny \big( \underline{u}  \big)} \\ B {\tiny \big( \underline{u}  \big)} & E {\tiny \big( \underline{u}  \big)} & H {\tiny \big( \underline{u}  \big)}  \\ C {\tiny \big( \underline{u}  \big) } &  F {\tiny \big( \underline{u}  \big) } & I {\tiny \big( \underline{u}  \big) }  
\end{bmatrix} \] \[ \overset{\bigotimes }{,}\begin{bmatrix}
 A {\tiny \big( \underline{u^{\prime}} \big)} & D {\tiny \big( \underline{u^{\prime}} \big)  }& G {\tiny \big( \underline{u^{\prime}} \big) } \\ B {\tiny \big( \underline{u^{\prime}}\big)} & E {\tiny \big( \underline{u^{\prime}} \big)} & H {\tiny \big( \underline{u^{\prime}} \big)}   \\ C {\tiny \big( \underline{u^{\prime}} \big)}  &  F {\tiny \big( \underline{u^{\prime}} \big)} & I {\tiny \big( \underline{u^{\prime}} \big) } 
\end{bmatrix} \bigg\} \text{, } 
\] }

\noindent for another representation of the product of three-dimensional L-operators,

\[
\begin{bmatrix}
 A {\tiny \big( \underline{u^{\prime}} \big)} & D {\tiny \big( \underline{u^{\prime}} \big)  }& G {\tiny \big( \underline{u^{\prime}} \big) } \\ B {\tiny \big( \underline{u^{\prime}}\big)} & E {\tiny \big( \underline{u^{\prime}} \big)} & H {\tiny \big( \underline{u^{\prime}} \big)}   \\ C {\tiny \big( \underline{u^{\prime}} \big)}  &  F {\tiny \big( \underline{u^{\prime}} \big)} & I {\tiny \big( \underline{u^{\prime}} \big) } 
\end{bmatrix}  \text{, } 
\]

\noindent in ${\tiny \underline{u^{\prime}}}$, which by the definition of Poisson bracket of the tensor product of the two reduced monodromy matrices, equals,

{\small  \begin{align*}
   \bigg\{    \bigg[  r^{3D}_{+} {\tiny \big( u_k - u^{\prime}_k , v_k - v^{\prime}_k , w_k - w^{\prime}_k \big)  }   \begin{bmatrix}
 A {\tiny \big( \underline{u} 
 \big)}  & D {\tiny \big( \underline{u} \big)}  & G {\tiny \big( \underline{u}  \big)} \\ B {\tiny \big( \underline{u}  \big)} & E {\tiny \big( \underline{u}  \big)} & H {\tiny \big( \underline{u}  \big)}  \\ C {\tiny \big( \underline{u}  \big) } &  F {\tiny \big( \underline{u}  \big) } & I {\tiny \big( \underline{u}  \big) }  
\end{bmatrix}   \bigg]   \otimes  \begin{bmatrix}
 A {\tiny \big( \underline{u^{\prime}} \big)} & D {\tiny \big( \underline{u^{\prime}} \big)  }& G {\tiny \big( \underline{u^{\prime}} \big) } \\ B {\tiny \big( \underline{u^{\prime}}\big)} & E {\tiny \big( \underline{u^{\prime}} \big)} & H {\tiny \big( \underline{u^{\prime}} \big)}   \\ C {\tiny \big( \underline{u^{\prime}} \big)}  &  F {\tiny \big( \underline{u^{\prime}} \big)} & I {\tiny \big( \underline{u^{\prime}} \big) } 
\end{bmatrix}  \bigg\} \\  - \bigg\{   \begin{bmatrix}
 A {\tiny \big( \underline{u} 
 \big)}  & D {\tiny \big( \underline{u} \big)}  & G {\tiny \big( \underline{u}  \big)} \\ B {\tiny \big( \underline{u}  \big)} & E {\tiny \big( \underline{u}  \big)} & H {\tiny \big( \underline{u}  \big)}  \\ C {\tiny \big( \underline{u}  \big) } &  F {\tiny \big( \underline{u}  \big) } & I {\tiny \big( \underline{u}  \big) }  
\end{bmatrix}  \otimes   \bigg[ \begin{bmatrix}
 A {\tiny \big( \underline{u^{\prime}} \big)} & D {\tiny \big( \underline{u^{\prime}} \big)  }& G {\tiny \big( \underline{u^{\prime}} \big) } \\ B {\tiny \big( \underline{u^{\prime}}\big)} & E {\tiny \big( \underline{u^{\prime}} \big)} & H {\tiny \big( \underline{u^{\prime}} \big)}   \\ C {\tiny \big( \underline{u^{\prime}} \big)}  &  F {\tiny \big( \underline{u^{\prime}} \big)} & I {\tiny \big( \underline{u^{\prime}} \big) } 
\end{bmatrix}  r^{3D}_{-} {\tiny \big( u_k - u^{\prime}_k , v_k - v^{\prime}_k , w_k - w^{\prime}_k \big) }          \bigg] \bigg\}  \text{, } 
\end{align*} } 

\noindent where the three-dimensional transfer matrix defined in the next section takes the form, for some vector ${\tiny \underline{M} \equiv \big( M_1 , M_2 \big)}$ over $\textbf{R}^2$, a negative real ${\tiny N}$, and natural ${\tiny j,k}$,

{\tiny  \begin{align*}
 \textbf{T}^{3D} \big( \underline{\lambda} \big)   \equiv    \underset{ N \longrightarrow +  \infty}{\underset{\underline{M} \longrightarrow + \infty}{\mathrm{lim}}} \mathrm{tr} \bigg\{     \overset{\underline{M}}{\underset{j=0}{\prod}}  \overset{0}{\underset{k=-N}{\prod}} \mathrm{exp} \big( \lambda_3 ( q^{-2} \xi^{s^j_k} ) \big)      \begin{bmatrix}     q^{D^j_k}       &    q^{-2} a^j_k q^{-D^j_k -D^j_{k+1}} \xi^{s-s^k_j}        &  a^j_k a^j_{k+1} q^{-D^j_k - 3D^j_{k+1}} \xi^{s - s^j_k - s^j_{k+1}}  \\ \big( a^j_k \big)^{\dagger} q^{D^j_k} \xi^{s^j_k} 
             &      q^{-D^j_k + D^j_{k+1}} - q^{-2} q^{D^j_k -D^j_{k+1}} \xi^{s}     &     - a^j_k q^{D^j_k - 3D^j_{k+1}} \xi^{s-s^j_k}  \\ 0  &    a^{\dagger}_j q^{D^j_k} \xi^{s^j_k} &  q^{-D^j_k} \\   \end{bmatrix}     \bigg\}  
\text{. }
\end{align*} }

\noindent The notion of weak integrability is given by the fact that,

{\small \[
 (\textit{Weak Integrability of the 20-vertex model}) \equiv  \text{ } 
\left\{\!\begin{array}{ll@{}>{{}}l} \big\{ A {\tiny \big( \underline{u} \big) } , A {\tiny \big( \underline{u^{\prime}} \big) } \big\}   \big/ \big\{ A {\tiny \big( u \big) } , A {\tiny \big( u^{\prime} \big) } \big\}   \longrightarrow + \infty   , \\ \big\{ B {\tiny \big( \underline{u} \big) } , B {\tiny \big( \underline{u^{\prime}} \big) } \big\}   \big/ \big\{ C  {\tiny \big( u \big) } , C {\tiny \big( u^{\prime} \big) } \big\}   \longrightarrow + \infty   ,  \\   \big\{ D {\tiny \big( \underline{u} \big) } , D {\tiny \big( \underline{u^{\prime}} \big) } \big\}   \big/ \big\{ B  {\tiny \big( u \big) } , B {\tiny \big( u^{\prime} \big) } \big\}   \longrightarrow + \infty       ,    \\   \big\{ E {\tiny \big( \underline{u} \big) } , E {\tiny \big( \underline{u^{\prime}}  \big) } \big\}   \big/ \big\{ D  {\tiny \big( u \big) } , D {\tiny \big( u^{\prime} \big) } \big\}   \longrightarrow + \infty     , \\    \\  \big\{ G {\tiny \big( \underline{u} \big)} , G {\tiny \big( \underline{u^{\prime}} \big) } \big\}  \big/ \mathcal{S}  \longrightarrow C {\tiny \big( G , \underline{u} , \underline{u^{\prime}} \big)      } \in {\tiny \big( 0 , + \infty )  }  , \\     \big\{ H {\tiny \big( \underline{u} \big)} , H {\tiny \big( \underline{u^{\prime}} \big) }  \big\}   \big/ \mathcal{S}    \longrightarrow C {\tiny \big( H , \underline{u} , \underline{u^{\prime}} \big)      } \in {\tiny \big( 0 , + \infty )  }       , \\         \big\{ I {\tiny \big( \underline{u} \big)} , I  {\tiny \big( \underline{u^{\prime}} \big) }  \big\}   \big/ \mathcal{S}   \longrightarrow C {\tiny \big( I  , \underline{u} , \underline{u^{\prime}} \big)      } \in {\tiny \big( 0 , + \infty )  }        , \\    
\end{array}\right.
\] } 

\noindent where ${\tiny \mathcal{S}} \equiv \mathrm{sup} \big\{   {\tiny \big\{ A {\tiny \big( u \big) } , A {\tiny \big( u^{\prime} \big) } \big\} , \big\{ B {\tiny \big( u \big) } , B  {\tiny \big( u^{\prime} \big) } \big\} , \big\{ C {\tiny \big( u \big) } , C  {\tiny \big( u^{\prime} \big) } \big\} , \big\{ D {\tiny \big( u \big) } , D {\tiny \big( u^{\prime} \big) } \big\}  }      \big\} $. The above rate of decay of the Poisson brackets from the block operators $A {\tiny \big( \underline{u} \big)}, \cdots, I {\tiny \big( \underline{u} \big)}$, as opposed to $A {\tiny \big( {u} \big)}, \cdots, D {\tiny \big( {u} \big)}$, is related to the existence of action-angle coordinates for which,

{\small \begin{align*}
        \big\{  \Phi^{3D} {\tiny \big( \underline{\lambda} \big) }  ,   \bar{\Phi^{3D} {\tiny \big( \underline{\lambda} \big) }  } \big\} = 0           \text{. } 
\end{align*} }

\noindent Moreover, in comparison to previous results under domain-wall boundary conditions for L-operators of the 6-vertex model, [24], the action of the Dynkin automorphism can be used to obtain other L-operators, and hence, approximations of the transfer matrix, through the following actions:

{\small

\begin{itemize}
\item[$\bullet$] \textit{Twisting the root map generator of the underlying Lie algebra}. The outer involution associated with the Dynkin automorphism on diagrams produces a twisted generator of the roots of the algebra produced by the action of the involution on the underlying Lie algebra.

\item[$\bullet$] \textit{Construction of the intertwining matrix}. Akin to intertwining operations that can be introduced for Solid-on-Solid models from the 20-vertex model, [42], the intertwining matrix $U$ for the ordinary 20-vertex model which one obtains from the involution of the Dynkin automorphism takes the form,

{\small

\begin{align*}
      \rho {\tiny \big(      \nu \big( X \big)   \big) } = {\tiny     U \rho \big( X \big) U^{-1}    }  , 
\end{align*}

}

\noindent for the action of the involution on the Lie algebra, ${\tiny \nu \big( X \big)}$.

\item[$\bullet$] \textit{Twisting of the L-operator}. L-operators, such as the one previously discussed, can be used to obtain other L-operators for the 20-vertex model which satisfy the statement of the Yang-Baxter equation from the Universal R-Matrix $\mathcal{R}$ (the equality $\mathcal{R}_{12} \mathcal{R}_{13} \mathcal{R}_{23} = \mathcal{R}_{23} \mathcal{R}_{13} \mathcal{R}_{12}$). The new L-operator takes the form,

{\small

\begin{align*}
   {\tiny  \widetilde{L} \big( \underline{\lambda} \big)   \equiv   \big\{   U_a    \otimes   U_q \big\}       L \big( \underline{\lambda} \big)    \big\{        U_a    \otimes   U_q   \big\}^{-1}          }     , 
\end{align*}

}

\noindent from the input L-operator $L$.

\end{itemize}

}

 \noindent From the relations above, there exists three-dimensional counterparts for $r_{a,+}$ and $r_{a,-}$, which are respectively given by,

{\small \begin{align*}
r^{3D}_{+} {\tiny \big( u_k - u^{\prime}_k , v_k - v^{\prime}_k , w_k - w^{\prime}_k \big)} \equiv  r^{3D}_{+ } \equiv     \underset{y \longrightarrow +\infty}{\mathrm{lim}}              \big\{  E^{3D,\mathrm{6V}} {\tiny \big(   \underline{u^{\prime}}  ,  v^{\prime}_k - v_k \big)}      \otimes  \big[                E^{3D,\mathrm{6V}} {\tiny \big(  \underline{u^{\prime}}  ,  v^{\prime}_k - v_k \big)}  \\ \times     r {\tiny \big(  u_k - u^{\prime}_k   , v_k - v^{\prime}_k  ,  w_k - w^{\prime}_k \big) }       \big] \big\}            \text{, } \\ \\ r^{3D}_{-} {\tiny \big( u_k - u^{\prime}_k , v_k - v^{\prime}_k , w_k - w^{\prime}_k \big)}  \equiv   r^{3D}_{-} \equiv      \underset{y \longrightarrow - \infty}{\mathrm{lim}}   \big\{ E^{3D,\mathrm{6V}} {\tiny \big(   \underline{u^{\prime}}  ,  v^{\prime}_k - v_k \big)  }     \otimes  \big[                E^{3D,\mathrm{6V}} {\tiny \big(  \underline{u^{\prime}}  ,  v^{\prime}_k - v_k \big)}   \\ \times    r {\tiny \big( u_k - u^{\prime}_k   , v_k - v^{\prime}_k  ,  w_k - w^{\prime}_k   \big)  }      \big] \big\}          \text{, } 
\end{align*} }

\noindent which are functions of three spectral parameters from differences, respectively, of $u_k$, $v_k$, and $w_k$. For,

{\small \begin{align*}
   E^{3D,\mathrm{6V}} {\tiny \big(   \underline{u^{\prime}}  ,  v^{\prime}_k - v_k ,  u^{\prime}_k - u_k  , w^{\prime}_k - w_k \big) }  \equiv    \mathrm{exp} \bigg\{  {\tiny  \big\{ 2i \big\}^{-1} \cdot \begin{bmatrix} 1 & 0 & 0 \\ 0 & -1 & 0 \\ 0 & 0 & 0 \end{bmatrix}  }  +  {\tiny \begin{bmatrix} 0 & 0 & \psi  \\ 0 & \bar{\psi}   & 0 \\ 0 & 0 & 0 \end{bmatrix}  }       \bigg\}        \text{, } 
\end{align*} }

\bigskip

\noindent given a solution $\psi$ to the Schrodinger's equation, and also for,

\begin{align*}
{\tiny  r^{3D}_{\pm} \big( u_k - u^{\prime}_k , v_k - v^{\prime}_k , w_k - w^{\prime}_k \big) }  \equiv    \underset{\underline{u^{\prime}} \longrightarrow \pm \infty}{\mathrm{lim}} \big\{     E^{3D,\mathrm{6V}} {\tiny \big(   \underline{u^{\prime}}  ,  v^{\prime}_k - v_k ,  u^{\prime}_k - u_k ,  w_k - w^{\prime}_k \big) } \\ \otimes     \big[ E^{3D,\mathrm{6V}} {\tiny \big(   \underline{u^{\prime}}  ,  v^{\prime}_k - v_k    , w_k - w^{\prime}_k \big) }   r^{3D} {\tiny \big( u_k - u^{\prime}_k , v_k - v^{\prime}_k ,  w^{\prime}_k - w_k  \big)  } \big]  \big\}        \text{. } 
\end{align*}

\noindent When manipulating Poisson brackets, one can make use of any of the following expressions, {[16]},

\begin{align*}
 \big\{ F ,  G \big\}   \equiv    i \cdot \int_{[-L,L]} \bigg\{          {\tiny {\delta F} \big/ {\delta \psi} \cdot {\delta G} \big/ {\delta \bar{\psi}} - {\delta F} \big/ {\delta \bar{\psi}} \cdot {\delta G} \big/ {\delta \psi }   }           \bigg\}   \mathrm{d} x      \text{, } 
\end{align*}

\noindent for functionals $F$ and $G$, while for the second case, the bracket takes the form,

{\small \begin{align*}
 \big\{  A \overset{\bigotimes }{,} B \big\} \equiv     i \cdot \int_{[-L,L]} \bigg\{   {\tiny {\delta A} \big/ {\delta \psi} \otimes  {\delta B} \big/ {\delta \bar{\psi}} - {\delta A} \big/ {\delta \bar{\psi}} \otimes   {\delta B}\big/ {\delta \psi }        }              \bigg\}     \mathrm{d} x     \text{, } 
\end{align*} }

\noindent for matrix functionals $A$ and $B$, and with ${\tiny \psi \equiv \psi \big( x \big)}$ and ${\tiny \bar{\psi} \equiv \bar{\psi} \big( x \big)}$ having support over ${\tiny \big[ - L , L \big]}$. For triangular ice, the Poisson bracket of the tensor product of two three-dimensional transfer matrices degenerates to the set of $81$ relations provided earlier in the section. Furthermore, as in the two-dimensional case, the following identities, originally introduced in seminal work on Hamiltonian methods, {[16]}, for the tensor product of the Poisson bracket of two transfer matrices, in three dimensions satisfies,

{\small \begin{align*}
      \big\{  T_{-} {\tiny \big( x , \underline{\lambda} \big)}  \overset{\bigotimes }{,} T_{-} {\tiny \big( x , \underline{\mu} \big)       }     \big\} = r {\tiny \big( \underline{\lambda} - \underline{\mu} \big) } T_{-} {\tiny \big( x , \underline{\lambda} \big) } \otimes  T_{-} {\tiny \big( x , \underline{\mu} \big)}  - T_{-} {\tiny \big( x , \underline{\lambda} \big) } \otimes  T_{-} {\tiny \big( x , \underline{\mu} \big)}  r_{-} {\tiny \big( \underline{\lambda} - \underline{\mu} \big)      }    \text{, } \\  \\    \big\{  T_{+} {\tiny \big( x , \underline{\lambda} \big)}  \overset{\bigotimes }{,} T_{+} {\tiny \big( x , \underline{\mu} \big)       }     \big\} = T_{+} {\tiny \big( x , \underline{\lambda} \big) } \otimes  T_{+} {\tiny \big( x , \underline{\mu} \big) }  r_{+} {\tiny \big( \underline{\lambda} - \underline{\mu} \big)}  - r {\tiny \big( \underline{\lambda} - \underline{\mu} \big)}  T_{+} {\tiny \big( x , \underline{\lambda} \big)}  \otimes  T_{+} {\tiny \big( x , \underline{\mu} \big) }   \text{, } 
\end{align*} }

\noindent also extend over to the three-dimensional case, in which,

{\small \begin{align*}
      \big\{  \textbf{T}_{-} {\tiny \big( x , \underline{\lambda} \big) } \overset{\bigotimes }{,} \textbf{T}_{-} {\tiny \big( x , \underline{\mu} \big)  }          \big\} = r^{3D} {\tiny \big( \underline{\lambda} - \underline{\mu} \big) } \textbf{T}_{-} {\tiny \big( x , \underline{\lambda} \big)} \otimes  \textbf{T}_{-} {\tiny \big( x , \underline{\mu} \big)}  - \textbf{T}_{-} {\tiny \big( x , \underline{\lambda} \big)} \otimes  \textbf{T}_{-} {\tiny \big( x , \underline{\mu} \big)}  r^{3D}_{-} {\tiny \big( \underline{\lambda} - \underline{\mu} \big)  }    \text{, } \\   \\   \big\{  \textbf{T}_{+} {\tiny \big( x , \underline{\lambda} \big) } \overset{\bigotimes }{,} \textbf{T}_{+} {\tiny \big( x , \underline{\mu} \big)   }         \big\} = \textbf{T}_{+} {\tiny \big( x , \underline{\lambda} \big) } \otimes  \textbf{T}_{+} {\tiny \big( x , \underline{\mu} \big) } r^{3D}_{+} {\tiny  \big( \underline{\lambda} - \underline{\mu} \big)}  - r^{3D} {\tiny \big( \underline{\lambda} - \underline{\mu}  \big)}  \textbf{T}_{+} {\tiny \big( x , \underline{\lambda} \big) } \otimes  \textbf{T}_{+} {\tiny \big( x , \underline{\mu} \big)    }     \text{, } 
\end{align*}}

\noindent where,

{\small \begin{align*}
 T_{\pm} {\tiny \big( x , \underline{\lambda} \big)}  = \underset{y \longrightarrow \pm \infty}{\mathrm{lim}} T {\tiny \big( x , y , \underline{\lambda} \big)}  E {\tiny \big( y , \underline{\lambda} \big) }    \text{, } 
\end{align*} } 

\noindent corresponds to the two-dimensional limit as ${\tiny y} \longrightarrow {\tiny \pm \infty}$, and where,

{\small \begin{align*}
\textbf{T}^{3D}_{\pm} {\tiny \big( x , \underline{\lambda} \big)}  \equiv  \textbf{T}_{\pm} {\tiny \big( x , \underline{\lambda} \big)}  = \underset{y \longrightarrow \pm \infty}{\mathrm{lim}} \textbf{T} {\tiny \big( x , y , \underline{\lambda} \big) } E^{3D} {\tiny \big( y , \underline{\lambda} \big) }    \text{. } 
\end{align*} }

\noindent corresponds to the three-dimensional limit as ${\tiny y} \longrightarrow {\tiny \pm \infty}$, for,

{\small \begin{align*}
  r^{3D}_{\pm} {\tiny \big( \underline{\lambda} - \underline{\mu} \big) } = \underset{x \longrightarrow + \infty}{\underset{y \longrightarrow \pm \infty}{\mathrm{lim}}} \big\{   E^{3D} {\tiny \big( x,y , \underline{\mu} - \underline{\lambda} \big) }   \otimes  \big[ E^{3D} {\tiny \big( x,y , \underline{\lambda} - \underline{\mu} \big)}  r^{3D} {\tiny \big( \underline{\lambda} - \underline{\mu} \big) }  \big] \big\}     \text{. } 
\end{align*} } 

\noindent Besides taking the limit of spatial variables of the three-dimensional transfer matrix to $+ \infty$, or to $- \infty$, several computations with L-operators, described in \textit{2.1}, imply,

{\small \begin{align*}
\underset{\underline{M} \longrightarrow + \infty}{\underset{N \longrightarrow + \infty}{\mathrm{lim}}}  \textbf{T}^{3D} {\tiny \big( \underline{M} , N , \underline{\lambda} \big)}  = \textbf{T}^{3D} {\tiny \big( \underline{\lambda} \big)  } \text{. } 
\end{align*} }

\noindent Relatedly, from the expression introduced for the three-dimensional quantum monodromy matrix, $T^{3D}_{a,b} \equiv T^{3D}$, in \textit{1.6}, one also has,

{\small \begin{align*}
\underset{\underline{M} \longrightarrow + \infty}{\underset{N \longrightarrow + \infty}{\mathrm{lim}}}  \textbf{T}^{3D}_{\underline{M},N} {\tiny \big( \underline{\lambda} \big)}  \equiv \underset{\underline{M} \longrightarrow + \infty}{\underset{N \longrightarrow + \infty}{\mathrm{lim}}}      \bigg\{     E^{3D,\mathrm{6V}} {\tiny \big(  \underline{M}  ,  v^{\prime}_k - v_k ,  u^{\prime}_k - u_k , w^{\prime}_k - w_k \big) }        \textbf{T}^{3D} {\tiny \big( \underline{M} , N  \big) }     E^{3D,\mathrm{6V}} {\tiny \big(  N ,  v^{\prime}_k}  \\{\tiny - v_k ,  u^{\prime}_k - u_k   , w^{\prime}_k - w_k   \big) } \bigg\}            \propto  \underset{\underline{M} \longrightarrow + \infty}{\underset{N \longrightarrow + \infty}{\mathrm{lim}}}  T^{3D}_a {\tiny \big( \underline{M} , N  \big) } = T^{3D}_a   \text{. } 
\end{align*} }

\noindent Finally, for action-angle variables in two dimensions which satisfy, ${\small  
        \big\{  \Phi {\tiny \big( \underline{\lambda} \big) }   ,   \bar{\Phi {\tiny \big( \underline{\lambda} \big) }  } \big\} = 0       \text{, } }$ in three dimensions, from computations to be performed with the Poisson bracket on the set of $81$ relations, one would hope to show the existence of three dimensional action-angle coordinates for which one would similarly have, ${\small \big\{  \Phi^{3D} {\tiny \big( \underline{\lambda} \big) }  ,   \bar{\Phi^{3D} {\tiny \big( \underline{\lambda} \big) }  } \big\} = 0           \text{. } }$.

\subsection{Statement of the Main result}

\noindent For the main result, an approximation for each of the $81$ Poisson brackets is obtained. To simplify the number of times that the bilinearity property of the Poisson bracket, the computations of each respective Poisson bracket are carried out for a simpler representation of the product of infinitely many terms of the transfer matrix.

\bigskip

\noindent \textbf{Main result} (\textit{three-dimensional Poisson structure}). Fix naturals $ i \neq i^{\prime\prime}$, and $i^{\prime} \neq i^{\prime\prime\prime}$. For a collection of strictly positive constants, $\big\{ {\tiny \big( C_j \big)_1 } \big\}_{1 \leq j \leq 81}$, $\big\{ {\tiny \big( C_k \big)_2 } \big\}_{1 \leq k \leq 81}$, and $\big\{ {\tiny \big( C_l \big)_3}  \big\}_{1 \leq l \leq 81}$, there exists constants $C_1 , \cdots , C_{81}$ satisfying,

{ \[
\left\{\!\begin{array}{ll@{}>{{}}l} \boxed{(1)}:      \big\{  A {\tiny \big( \underline{u} \big)  }       , A {\tiny \big( \underline{u^{\prime}} \big) }   \big\} \approx  C_1 \equiv  \frac{                     (C_{1})_1   [ \underline{A ( \underline{u} )} ]  }{i^{\prime\prime} - i }  +      \frac{(C_{1})_2 [ \mathcal{T}_{(1,1))} ]}{i^{\prime\prime\prime} - i^{\prime}  }  + ( C_{1} )_3  
\text{ , } \\  \vdots  \\ \boxed{(81)}: \big\{ I {\tiny \big( \underline{u} \big) } , I {\tiny \big( \underline{u^{\prime}} \big) } \big\} \approx  C_{81} \equiv               \frac{                   ( C_{81} )_1 [ \underline{I ( \underline{u} )}] }{i^{\prime\prime} - i }  +      \frac{  ( C_{81} )_2 [ \mathcal{T}_{(3,3)} ] }{i^{\prime\prime\prime} - i^{\prime}  }   + ( C_{81} )_3   \text{ . }
\end{array}\right.
\]}

\noindent The above result is the mostly closely related to previous results in the literature which characterized integrability of inhomogeneous limit shapes for the 6-vertex model. For the sake of completeness we provide a statement of a few of the main results from this work, [24], below. The action-angle coordinates underlying the notion of Poisson commutativity of the 6-vertex Hamiltonian flow are related to the action-angle coordinates seminally introduced in [16].

\bigskip

\noindent \textbf{Theorem} (\textbf{Theorem} \textit{4.1}, [24]: \textit{the Hamiltonian flow for the inhomogeneous 6-vertex model under domain-wall boundary conditions constitutes a Poisson commuting family}). The Hamiltonian flow defined through the object,

{\small

\begin{align*}
        H_u \big( \pi \big( x , y \big) , h \big( x , y \big) \big) \equiv H_u \big( \pi , h \big) = {\int}_{[0,L]}   \mathcal{H}_{u - v ( x) } \big[  \partial_x h \big( x \big)  ,    \pi \big( x \big)   \big]  \mathrm{d} x    \text{, } 
\end{align*}

}

\noindent forms a Poisson commutative family, in which $ \big\{ {\tiny H_u , H_w } \big\} = 0$ for any $u,w$.

\bigskip 

\noindent \textbf{Claim} (\textit{Chapter} \textbf{I}, \textit{Section} \textbf{5}, [16]: \textit{the two-dimensional transfer matrix can be expressed in terms of a column span of two block operators}). Fix a spectral parameter $\lambda$. The finite-dimensional representation for the transfer matrix $T {\tiny \big( \lambda \big)}$,

{\small

\begin{align*}
   \begin{bmatrix} a {\tiny \big( \lambda \big)} &     \epsilon \cdot \tilde{b} {\tiny \big( \lambda \big)}       \\ b {\tiny \big( \lambda \big)}  & \tilde{a} {\tiny \big( \lambda \big)} 
     \end{bmatrix} .
\end{align*}

}

\noindent The above transition coefficients, ${\tiny a}$ and ${\tiny b}$, satisfy the condition ${\tiny \big| a \big( \lambda \big)  \big|^2 - \epsilon \big| b \big( \lambda \big) \big|^2 }$, for,

{\small

\begin{align*}
 \underset{L \longrightarrow + \infty}{\mathrm{lim}} {\tiny \mathrm{exp} \big[    i \lambda L      \big] \cdot a_L \big( \lambda \big) = a \big( \lambda \big) }  , \\    \underset{L \longrightarrow + \infty}{\mathrm{lim}} {\tiny \mathrm{exp} \big[    i \lambda L      \big] \cdot b_L \big( \lambda \big)      =  b \big( \lambda \big)  } .  
\end{align*}

}

\noindent \textbf{Claim} (\textit{Chapter} \textbf{III}, \textit{Section} \textbf{6}, [16]: \textit{the tensor product of the Poisson bracket for two transfer matrices decomposes into a collection of 16 Poisson brackets}). Fix the spectral parameters $\lambda \neq \mu$. The Poisson bracket ${\tiny \big\{          T \big( \lambda \big) \overset{\otimes}{,}  T \big( \mu \big))      \big\}  }$, from the decomposition,

{\small

\begin{align*}
   {\tiny   r_{+} \big( \lambda - \mu \big) T \big( \lambda \big) \otimes T \big( \mu \big) - T \big( \lambda \big) \otimes  T \big( \mu \big) r_{-} \big( \lambda - \mu \big)    }     ,
\end{align*}

}

\noindent where,

{\small

\begin{align*}
    r_{\pm } \equiv  r_{a,\pm}      {\tiny \big( v_k - v^{\prime}_k  \big) }  =       \underset{y \longrightarrow \pm  \infty}{\mathrm{lim}}  \big\{ E^{\mathrm{6V}} {\tiny \big(   u^{\prime},  v^{\prime}_k - v_k  \big)}  \otimes  \big[                E^{\mathrm{6V}} {\tiny \big(   u^{\prime},  v^{\prime}_k - v_k   \big)  }     r_a {\tiny \big(  v_k - v^{\prime}_k  \big)   }         \big]  \big\}  \text{, }      \end{align*}

 \begin{align*} 
       E^{\mathrm{6V}} {\tiny \big( v  - v_k  , x \big) } = \mathrm{exp} \big[             \mathrm{coth} {\tiny \big( \frac{\eta}{2} + i \alpha_j - v_k \big)    }         \big]      ,
\end{align*}

}

\noindent is equivalent to the collection,

{\small

\begin{align*}
 {\tiny \big\{ a \big( \lambda \big) , a \big( \mu \big) \big\} } , \\  {\tiny   \big\{ a \big( \lambda \big)  ,  \tilde{a} \big( \mu \big)     \big\}   } , \\  {\tiny \big\{         a \big( \lambda \big) , b \big( \mu \big) \big\}  } , \\  {\tiny \big\{  a \big( \lambda \big) , \tilde{b} \big( \mu \big) \big\} } , \\  {\tiny   \big\{ b \big( \lambda \big) , b \big( \mu \big) \big\}    } , \\  {\tiny   \big\{ b \big( \lambda \big) , \tilde{b} \big( \mu \big) \big\}    } ,
\end{align*}

}

\noindent of $6$ Poisson brackets. The above collection of relations is equivalent to the collection of $16$ Poisson brackets given block operators for the transfer matrix of the 6-vertex model.

\section{Three-dimensional transfer matrix}

The transfer matrix defined in the previous section will be analyzed from sets of relations obtained from products of three-dimensional L-operators.

\subsection{Recursive set of relations for the the transfer matrix when holding one spectral parameter constant}

\noindent For $\underline{M},N$ taken sufficiently large,

{\tiny  \begin{align*}
     \mathrm{tr} \bigg\{     \overset{\underline{M}}{\underset{\underline{j}=0}{\prod}}  \text{ }  \overset{0}{\underset{i=-N}{\prod}} \mathrm{exp} \big( \lambda_3 ( q^{-2} \xi^{s_i} ) \big)     \begin{bmatrix}     q^{D_i}       &    q^{-2} a_i q^{-D_i-D_j} \xi^{s-s_i}        &   a_i a_{j} q^{-D_i - 3D_j} \xi^{s - s_i - s_j}  \\ a^{\dagger}_i q^{D_i} \xi^{s_i} 
             &      q^{-D_i + D_j} - q^{-2} q^{D_i -D_j} \xi^{s}     &     - a_j q^{D_i - 3D_j} \xi^{s-s_j}  \\ 0  &    a^{\dagger}_j q^{D_j} \xi^{s_j} &  q^{-D_j} \\    \end{bmatrix}        \bigg\}    \text{, } 
\end{align*} }

\noindent we decompose the above trace operation through products of L-operators taking the form,

{\small \[
 \begin{bmatrix}
   q^{D_0}     &    q^{-2} a_0 q^{-D_0 - D_j}   &       a_0 a_j q^{D_0 - 3 D_j} \xi^{s-s_0-s_j }  \\   a^{\dagger}_0 q^{D_0} \xi^{s_0 }   &              q^{-D_0 - D_j} - q^{-2} q^{-3D_0 - D_j} \xi^s         &  - a_j q^{D_0 - 3 D_j}   \xi^{s-_j}                  \\  0     &  a^{\dagger}_j q^{D_j} \xi^{s_0}     &            q^{-D_j}      \\ 
\end{bmatrix}       \text{, } 
\] }

\noindent which consists of contributions from the complex-valued mapping $\xi$ into the unital associative algebra, in addition to contributions from raising the $q$ parameter to powers of differential operators.

\bigskip

\noindent As a matter of notation, introduce the three-dimensional limit of products of L-operators in the asymptotic limit as,

{\small   \begin{align*}
\underset{N \longrightarrow + \infty}{\mathrm{lim}}  { \underset{ 0 \leq i \leq  N} {\prod}\begin{bmatrix}
  \textbf{1}^i  & \textbf{4}^i & \textbf{7}^i     \\   \textbf{2}^i & \textbf{5}^i & \textbf{8}^i   \\ \textbf{3}^i & \textbf{6}^i & \textbf{9}^i 
    \end{bmatrix} \propto \begin{bmatrix}
 A {\tiny \big( \underline{u^{\prime}} \big)}  & D {\tiny \big(\underline{u^{\prime}} \big)}  & G {\tiny \big(\underline{u^{\prime}}\big)}  \\ B {\tiny \big( \underline{u^{\prime}}\big) } & E {\tiny \big( \underline{u^{\prime}} \big)}  & H {\tiny \big(\underline{u^{\prime}} \big) }  \\ C {\tiny \big(\underline{u^{\prime}} \big)}   &  F {\tiny \big( \underline{u^{\prime}}\big)} & I {\tiny \big( \underline{u^{\prime}} \big) }  
\end{bmatrix} }  \propto \bigg\{    { \begin{bmatrix}
 A {\tiny \big( \underline{u} \big) } & D {\tiny \big( \underline{u} \big)}   & G {\tiny \big( \underline{u} \big)}  \\ B {\tiny \big( \underline{u} \big)}  & E {\tiny \big( \underline{u} \big)}  & H {\tiny \big( \underline{u} \big) } \\ C {\tiny \big( \underline{u} \big)}   &  F {\tiny \big( \underline{u}  \big)}  & I {\tiny \big( \underline{u} \big)}  
\end{bmatrix}  } \\ \overset{\bigotimes }{,} { \begin{bmatrix}
 A {\tiny \big( \underline{u^{\prime}} \big)}  & D {\tiny \big( \underline{u^{\prime}} \big)}  & G {\tiny \big( \underline{u^{\prime}} \big)} \\ B {\tiny \big(\underline{u^{\prime}} \big)} & E {\tiny \big( \underline{u^{\prime}} \big)}  & H {\tiny  \big( \underline{u^{\prime}} \big) } \\ C {\tiny \big( \underline{u^{\prime}} \big)}   &  F \big( \underline{u^{\prime}} \big) & I \big( \underline{u^{\prime}} \big) 
\end{bmatrix} } \bigg\} \end{align*}

 \begin{align*} \propto  \bigg[ \underset{N \longrightarrow + \infty}{\mathrm{lim}} \bigg\{    \begin{bmatrix}
 A {\tiny \big( \underline{u} \big)}  & D {\tiny \big( \underline{u} \big)}   & G {\tiny \big( \underline{u} \big)}  \\ B {\tiny \big( \underline{u} \big)}  & E {\tiny \big( \underline{u} \big)}  & H {\tiny \big( \underline{u} \big) }  \\ C {\tiny \big( \underline{u} \big)}   &  F {\tiny \big( \underline{u}  \big)} & I {\tiny \big( \underline{u} \big) } 
\end{bmatrix} \overset{\bigotimes }{,}\begin{bmatrix}
 A_N {\tiny \big( \underline{u^{\prime}} \big)}  & D_N {\tiny \big( \underline{u^{\prime}} \big) }  & G_N {\tiny \big( \underline{u^{\prime}} \big) } \\ B_N {\tiny \big(\underline{u^{\prime}} \big)}  & E_N {\tiny \big( \underline{u^{\prime}} \big)}  & H_N {\tiny \big( \underline{u^{\prime}} \big)}   \\ C_N {\tiny \big( \underline{u^{\prime}} \big)}   &  F_N {\tiny \big( \underline{u^{\prime}} \big)}  & I_N {\tiny \big( \underline{u^{\prime}} \big)}  
\end{bmatrix} \bigg\} \bigg]  \text{. }  
\end{align*} }

\noindent To quantify asymptotic expansions of the transfer matrix, one performs computations of,

{\small \begin{align*}
\underset{\underline{M} \longrightarrow + \infty}{\mathrm{lim}} \bigg\{  \underset{ 0 \leq \underline{j} \leq \underline{M}} {\prod}\begin{bmatrix}
  \textbf{1}^{\underline{j}} & \textbf{4}^{\underline{j}} & \textbf{7}^{\underline{j}}     \\   \textbf{2}^{\underline{j}} & \textbf{5}^{\underline{j}} & \textbf{8}^{\underline{j}}   \\ \textbf{3}^{\underline{j}} & \textbf{6}^{\underline{j}} & \textbf{9}^{\underline{j}} 
    \end{bmatrix} \bigg\} 
    \text{, }      \\ 
\underset{N \longrightarrow + \infty}{\mathrm{lim}} \bigg\{  \underset{\underline{M} \longrightarrow + \infty}{\mathrm{lim}}  \underset{ 0 \leq \underline{j} \leq \underline{M}} {\prod}  \bigg\{ \underset{ - N \leq i \leq 0} {\prod} 
 \begin{bmatrix}
  \textbf{1}^{i,\underline{j}} & \textbf{4}^{i,\underline{j}} & \textbf{7}^{i,\underline{j}}     \\   \textbf{2}^{i,\underline{j}} & \textbf{5}^{i,\underline{j}} & \textbf{8}^{i,\underline{j}}   \\ \textbf{3}^{i,\underline{j}} & \textbf{6}^{i,\underline{j}} & \textbf{9}^{i,\underline{j}} 
    \end{bmatrix} \bigg\} \bigg\} 
    \text{, }      \end{align*} }

\noindent each of which correspond to the weak infinite volume expansion of $\textbf{T}$ for indices, $i,\underline{j}$, of the finite volume representation of the transfer matrix.

\subsection{Three-dimensional product representations of L-operators}

\noindent We state the main result which demonstrates how the block operators $A {\tiny \big( \underline{u} \big)}, B {\tiny \big( \underline{u} \big)}, C {\tiny \big( \underline{u} \big)}$ can be obtained from the product of four L-operators.

\bigskip

\noindent \textbf{Lemma} \textit{P1}  (\textit{product representation of L-operators up to the fourth order}). The product of four L-operators is spanned by the union of three subspaces,

{\small \begin{align*}
    \begin{bmatrix}
      A {\tiny \big( \underline{u}  \big) }    \\ B {\tiny \big( \underline{u} \big) }  \\ C {\tiny \big( \underline{u} \big)
   }  \end{bmatrix} \equiv \begin{bmatrix}
      1    \\    2    \\ 3
    \end{bmatrix}
\end{align*} }

\noindent for,

{\tiny  \begin{align*}
 1 \equiv  \bigg[   \underline{\mathcal{T}^1_{(1,1)}} + \underline{\mathcal{T}^2_{(1,1)}}  + \underline{\mathcal{T}^3_{(1,1)}}  + \underline{\mathcal{T}^4_{(1,1)}}  + \underline{\mathcal{T}^5_{(1,1)}} + \underline{\mathcal{T}^6_{(1,1)}}      \bigg]  \bigg[  \underline{A_1 \big( \underline{u} \big) } 
        + \underline{A_2 \big( \underline{u} \big) }  + \underline{A_3 \big( \underline{u} \big) }  + \underline{A_4  \big( \underline{u} \big) }  + \underline{A_5 \big( \underline{u} \big) }  \bigg] \\ \times \bigg[     \underset{i \in \textbf{N} :  |  i    | \leq  4 \lceil \frac{m-4}{3} \rceil }{\prod}   \underline{A_{1,i} \big( \underline{u}\big) }   + \underset{i \in \textbf{N} :  |  i    | \leq  4 \lceil \frac{m-4}{3} \rceil }{\prod}   \underline{A_{2,i} \big( \underline{u}\big) }           + \underset{i \in \textbf{N} :  |  i    | \leq  4 \lceil \frac{m-4}{3} \rceil }{\prod}   \underline{A_{3,i} \big( \underline{u}\big) }    + \underset{i \in \textbf{N} :  |  i    | \leq  4 \lceil \frac{m-4}{3} \rceil }{\prod}   \underline{A_{4,i} \big( \underline{u}\big) } \\  + \underset{i \in \textbf{N} :  |  i    | \leq  4 \lceil \frac{m-4}{3} \rceil }{\prod}   \underline{A_{5,i} \big( \underline{u}\big) }  \bigg]    \text{, } \\ \\     2 \equiv   \bigg[ \underline{\mathcal{T}^1_{(1,2)} }+            \underline{\mathcal{T}^2_{(1,2)}}   + \underline{\mathcal{T}^3_{(1,2)}}  + \underline{\mathcal{T}^4_{(1,2)}}  \bigg]  \bigg[    \underline{B_1 \big( \underline{u} \big) }          + \underline{B_2 \big( \underline{u} \big) }   + \underline{B_3 \big( \underline{u} \big) }    + \underline{B_4 \big( \underline{u} \big) }   \bigg]                     \bigg[ \underset{i \in \textbf{N} :  |  i    | \leq  4 \lceil \frac{m-4}{3} \rceil }{\prod}   \underline{B_{1,i} \big( \underline{u}\big) }  \\ + \underset{i \in \textbf{N} :  |  i    | \leq  4 \lceil \frac{m-4}{3} \rceil }{\prod}   \underline{B_{2,i} \big( \underline{u}\big) }         +  \underset{i \in \textbf{N} :  |  i    | \leq  4 \lceil \frac{m-4}{3} \rceil }{\prod}   \underline{B_{3,i} \big( \underline{u}\big) }  + \underset{i \in \textbf{N} :  |  i    | \leq  4 \lceil \frac{m-4}{3} \rceil }{\prod}   \underline{B_{4,i} \big( \underline{u}\big) }    \bigg]     \text{, } \\ \\ 3 \equiv     \bigg[  \underline{\mathcal{T}^1_{(1,3)}} + \underline{\mathcal{T}^2_{(1,3)}}  + \underline{\mathcal{T}^3_{(1,3)}}  + \underline{\mathcal{T}^4_{(1,3)}}    \bigg]  \bigg[ \underline{C_1 \big( \underline{u} \big) }         +    \underline{C_2 \big( \underline{u} \big) }    + \underline{C_3 \big( \underline{u} \big) }     + \underline{C_4 \big( \underline{u} \big) }                 \bigg]   \bigg[ \underset{i \in \textbf{N} :  |  i    | \leq  4 \lceil \frac{m-4}{3} \rceil }{\prod}   \underline{C_{1,i} \big( \underline{u}\big) } \\ + \underset{i \in \textbf{N} :  |  i    | \leq  4 \lceil \frac{m-4}{3} \rceil }{\prod}   \underline{C_{2,i} \big( \underline{u}\big) }  + \underset{i \in \textbf{N} :  |  i    | \leq  4 \lceil \frac{m-4}{3} \rceil }{\prod}   \underline{C_{3,i} \big( \underline{u}\big) } + \underset{i \in \textbf{N} :  |  i    | \leq  4 \lceil \frac{m-4}{3} \rceil }{\prod}   \underline{C_{4,i} \big( \underline{u}\big) }  \bigg]          \text{, }
\end{align*} }

 \noindent corresponding to the product,

{\small \[ \bigg\{ \begin{bmatrix}
  \textbf{1}^{1,\underline{j}} & \textbf{4}^{1,\underline{j}} & \textbf{7}^{1,\underline{j}}     \\   \textbf{2}^{1,\underline{j}} & \textbf{5}^{1,\underline{j}} & \textbf{8}^{1,\underline{j}}   \\ \textbf{3}^{1,\underline{j}} & \textbf{6}^{1,\underline{j}} & \textbf{9}^{1,\underline{j}} 
    \end{bmatrix}  \begin{bmatrix}
  \textbf{1}^{1,\underline{j}+1} & \textbf{4}^{1,\underline{j}+1} & \textbf{7}^{1,\underline{j}+1}     \\   \textbf{2}^{1,\underline{j}+1} & \textbf{5}^{1,\underline{j}+1} & \textbf{8}^{1,\underline{j}+1}   \\ \textbf{3}^{1,\underline{j}+1} & \textbf{6}^{1,\underline{j}+1} & \textbf{9}^{1,\underline{j}+1} 
    \end{bmatrix}  \begin{bmatrix}
  \textbf{1}^{1,\underline{j}+2} & \textbf{4}^{1,\underline{j}+2} & \textbf{7}^{1,\underline{j}+2}     \\   \textbf{2}^{1,\underline{j}+2} & \textbf{5}^{1,\underline{j}+2} & \textbf{8}^{1,\underline{j}+2}   \\ \textbf{3}^{1,\underline{j}+2} & \textbf{6}^{1,\underline{j}+2} & \textbf{9}^{1,\underline{j}+2} 
    \end{bmatrix}  \bigg\} \begin{bmatrix}
  \textbf{1}^{1,\underline{j}+3} & \textbf{4}^{1,\underline{j}+3} & \textbf{7}^{1,\underline{j}+3}     \\   \textbf{2}^{1,\underline{j}+3} & \textbf{5}^{1,\underline{j}+3} & \textbf{8}^{1,\underline{j}+3}   \\ \textbf{3}^{1,\underline{j}+3} & \textbf{6}^{1,\underline{j}+3} & \textbf{9}^{1,\underline{j}+3} 
    \end{bmatrix} \times \cdots   \text{. } \] }

\bigskip

\noindent \textbf{Lemma} \textit{P2} and \textbf{Lemma} \textit{P3}, which provide a similarly defined system of relations for the span of the two remaining columns of the transfer matrix, are provided in the Appendix. We establish that the product representation of block operators over the first column of the transfer matrix holds through the following results.

\bigskip 

\noindent \textbf{Lemma} \textit{1} (\textit{the product representation when varying one spectral parameter of the twenty-vertex model, the base case induction step for lower order terms of the three-dimensional expansion}). Suppose that \textbf{Lemma} \textit{20V-CP-1} and \textbf{Lemma} \textit{20V-CP-2} hold. One has that the product,

{\tiny \[
    \begin{bmatrix}     q^{D^j_2}              &  q^{-2} a^j_3 q^{-D^j_3 - D^j_1}     \xi^{s-s^j_3}        &     a^j_3 a^j_1 q^{-D^j_3 - 3 D^j_1} \xi^{s-s^j_3 - s^j_1}         \\    \big( a^j_3  \big)^{\dagger} q^{D^j_3} \xi^{s^j_3}        &   q^{-D^j_3 + D^j_1} - q^{-2} q^{D_3 - D_1} \xi^s    &       - a_1 q^{D_3 - 3 D_1} \xi^{s-s_3 - s_1}  \\   0  &    a^{\dagger}_1 q^{D_3} \xi^{s_3}   &    - q^{D_1}    \\      \end{bmatrix} \begin{bmatrix}     q^{D_3}     &    q^{-2} a_3 q^{-D_3 - D_3}  \xi^{s-s_3}    &    a^2_3 q^{-D_3 - 3 D_3 }    \\             a^{\dagger}_3 q^{D_3} \xi^{s_3}           &   q^{-D_3 + D_2} - q^{-2} q^{D_3 - D_2}      &    -a_3 q^{D_3 - D_3} \xi^{s-s_3}  \\ 0   &     a^{\dagger}_3 q^{D_3} \xi^{s_3}     &   q^{-D_3}        \end{bmatrix} \text{, }     \]}

\noindent can be expressed in terms of the union of the span of the three subspaces,

{\tiny \[
\begin{bmatrix}
 q^{D^j_1 + D^j_2} + \big( a^j_1 \big)^{\dagger} q^{D^j_1 } \xi^{s^j_1} q^{-2} a^j_2 q^{-D^j_2 - D^j_3 } \xi^{s-s^j_2}   \\ \\   q^{D^j_1} \big( a^j_2 \big)^{\dagger} q^{D^j_2} \xi^{s^j_2} + \big( a^j_1 \big)^{\dagger}   q^{D^j_1} \xi^{s^j_1} q^{-D^j_2 + D^j_3}  - \big( a^j_1 \big)^{\dagger} q^{D^j_1 } \xi^{s^j_1} q^{-2} q^{D^j_2 - 3 d^j_3 } \xi^s           \\   \\   \big( a^j_1 \big)^{\dagger} q^{D^j_1} \xi^{s^j_1} \big( a^j_3 \big)^{\dagger} q^{D^j_3 } \xi^{s^j_3 }  \end{bmatrix} \text{, } 
\]

\begin{align*}\begin{bmatrix}
      q^{-2} a^j_1 q^{-D^j_1 - D^j_2} \xi^{s-s^j_1} q^{D^j_2} + q^{-D^j_1 + D^j_2} q^{-2} a^j_2 q^{-D^j_2 - D^j_3 } \xi^{s-s^j_2} - q^{-2} q^{D^j_1 - 3 D^j_2 } \xi^s q^{-2} a^j_2 q^{-D^j_2 - D^j_3} \xi^{s-s^j_2}  + \big( a^j_2 \big)^{\dagger} \\ \times q^{D^j_2} \xi^{s^j_2} a^j_2 a^j_3  q^{-D^j_2 - 3 D^j_3} \xi^{s-s^j_2 - s^j_3}        \\   \\ q^{-2} a^j_1 q^{-D^j_1 - D^j_2} \xi^{s-s^j_1} \big( a^j_2 \big)^{\dagger} q^{D^j_2} \xi^{s^j_2}   + q^{-D^j_1 + D^j_2 } q^{-D^j_2 + D^j_3} - q^{-2} q^{D^j_1 - 3 D^j_2 } \xi^s  q^{-D^j_2 + D^j_3} - q^{-D^j_1 + D^j_2}  q^{-2} q^{D^j_2 - 3 D^j_3} \xi^s  \\ + q^{-2} q^{D^j_1 - 3 D^j_2} \xi^s q^{-2} q^{D^j_2 - 3 D^j_3} \xi^s   +  \big( a^j_2 \big)^{\dagger} q^{D^j_2} \xi^{s^j_2} \big( - a^j_3 \big) q^{D^j_3 - 3 D^j_4} \xi^{s-s^j_3}         \\      \\        q^{-2} a^j_1 q^{-D^j_1 - D^j_2} \xi^{s-s^j_1} \big( a^j_2 \big)^{\dagger} q^{D^j_2} \xi^{s^j_2}   + q^{-D^j_1 + D^j_2 } q^{-D^j_2 + D^j_3} - q^{-2} q^{D^j_1 - 3 D^j_2 } \xi^s  q^{-D^j_2 + D^j_3} - q^{-D^j_1 + D^j_2}  q^{-2} q^{D^j_2 - 3 D^j_3} \xi^s  \\ + q^{-2} q^{D^j_1 - 3 D^j_2} \xi^s q^{-2} q^{D^j_2 - 3 D^j_3} \xi^s   +  \big( a^j_2 \big)^{\dagger} q^{D^j_2} \xi^{s^j_2} \big( - a^j_3 \big) q^{D^j_3 - 3 D^j_4} \xi^{s-s^j_3}      
\end{bmatrix}\end{align*}

\begin{align*}\begin{bmatrix}
              a^j_1 a^j_2 q^{-D^j_1 - 3 D^j_2} \xi^{s-s^j_1 - s^j_2} q^{D^j_2} - a^j_2 q^{D^j_2 - 3 D^j_3} \xi^{s-s^j_2} q^{-2} a^j_2 q^{-D^j_2 - D^j_3} \xi^{s-s^j_2} + q^{-D^j_2} a^j_2 a^j_3 q^{-D^j_2 - 3 D^j_2} \xi^{s-s^j_2 - s^j_3}       \\      \\      a^j_1 a^j_2 q^{-D^j_1 - 3 D^j_2} \xi^{s-s^j_1 - s^j_2} \big( a^j_2 \big)^{\dagger} q^{D^j_2} \xi^{s^j_2} - a^j_2 q^{D^j_2 - 3 D^j_3} \xi^{s-s^j_2} \big(  q^{-D^j_2 + D^j_3} -  q^{-2 } q^{D^j_2 - 3 D^j_3} \xi^s \big)  - q^{-D^j_2}  a^j_3 q^{D^j_3 - 3 D^j_4} \xi^{s-s^j_3}  \\   \\      - a^j_2 q^{D^j_2 - 3 D^j_3} \xi^{s-s^j_2} \big( a^j_3 \big)^{\dagger} q^{D^j_3} \xi^{s^j_3} + q^{-D^j_2 - D^j_3}       
\end{bmatrix} \end{align*} }

\noindent \textit{Proof of Lemma 1}. To characterize lower order terms of the infinite weak volume limit of the three-dimensional matrix, consider products of the type described before, namely,

{\small \begin{align*}
\mathcal{T}_1 \equiv   \begin{bmatrix}  q^{D^j_1 }  &   q^{-2} a^j_1  q^{-D^j_1 - D^j_2 }    &     a^j_1 a^j_2  q^{D^j_1 - 3 D^j_2  }  \xi^{s-s^j_1  - s^j_2  } \\  \big( a^j_1 \big)^{\dagger} q^{D^j_1}  \xi^{s^j_1 }    &  q^{-D^j_1  - D^j_2 } - q^{-2} q^{-3 D^j_1 - D^j_2  } \xi^s   &  - a^j_1  q^{D^j_1 - 3 D^j_2  } \xi^{s-s^j_1 }  \\   0   &  \big( a^j_1 \big)^{\dagger} q^{D^j_1 } \xi^{s^j_1 }  & q^{-D^j_1  }       \\     \end{bmatrix}\text{, }    \end{align*} }

\noindent with,

{\small \begin{align*}
 \mathcal{T}_2 \equiv  \begin{bmatrix}  q^{D^j_2 }  &   q^{-2} a^j_2  q^{-D^j_2 - D^j_3 }    &     a^j_2 a^j_3  q^{D^j_2 - 3 D^j_3  }  \xi^{s-s^j_2  - s^j_3  } \\  \big( a^j_2 \big)^{\dagger} q^{D^j_2}  \xi^{s^j_2 }    &  q^{-D^j_2  - D^j_3 } - q^{-2} q^{-3 D^j_2 - D^j_3  } \xi^s   &  - a^j_2  q^{D^j_2 - 3 D^j_3  } \xi^{s-s^j_2 }  \\   0   &  \big( a^j_2 \big)^{\dagger} q^{D^j_2 } \xi^{s^j_2 }  & q^{-D^j_2  }       \\     \end{bmatrix}   \text{, } \end{align*}  }

\noindent while holding the $j$ superscript of all terms in the L-operator constant (with the expressions obtained from the system below, an additional set of relations for the $j$ superscript will be obtained with terms from varying lower order terms of the $k$ subscript). To determine which terms appear in higher order terms in matrix products of L-operators (such as from the configurations of L-operators depicted in $\textbf{Figure}$ $\textit{1}$ and in $\textbf{Figure}$ $\textit{2}$), from expressions of the form,

\begin{figure}[H]
\begin{align*}
\includegraphics[width=0.5\columnwidth]{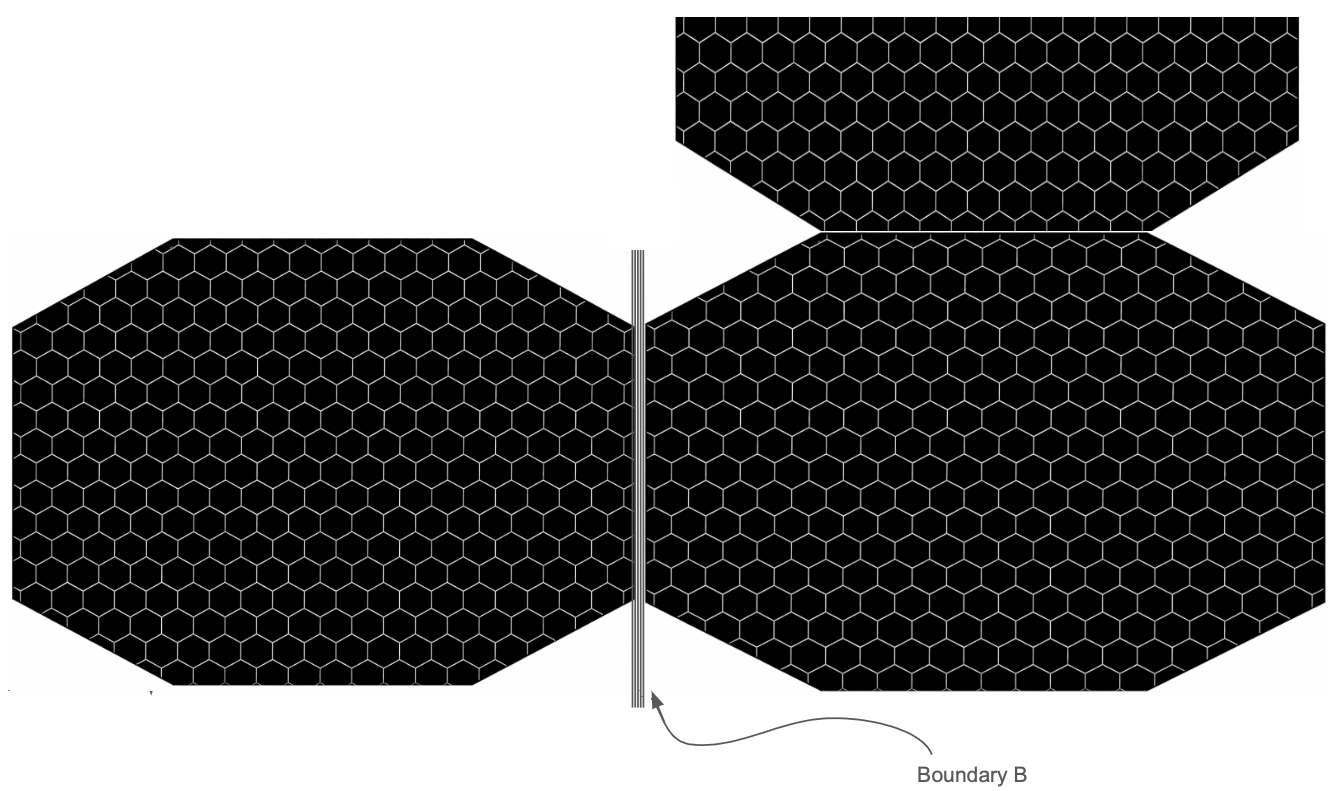}
\end{align*}
\caption{A depiction of three hexagons colored in black. When computing the entries of products of L-operators, one enforces boundary conditions that are propagated across the boundary $B$ between the rightmost and leftmost hexagons, as well as across another boundary between the topmost and leftmost hexagons.}
\end{figure}

{\small \begin{align*}
    \underset{ 0 \leq j \leq \underline{M}}{\prod}  \bigg\{ \begin{bmatrix}
 q^{D^j_1 } & q^{-2} a^j_1  q^{-D^j_1 - D^j_2} \xi^{s-s^j_1} & a^j_1 a^j_2 q^{-D^j_1 - 3 D^j_2} \xi^{s-s^j_1 - s^j_2} \\ \big( a^j_1 \big)^{\dagger} q^{D^j_1} \xi^{s^j_1}  & q^{-D^j_1 + D^j_2} - q^{-2} q^{D^j_1 - 3 D^j_2} \xi^s & -a^j_2 q^{D^j_2 - 3 D^j_3} \xi^{s-s^j_2} \\ 0 & \big( a^j_2 \big)^{\dagger} q^{D^j_2 } \xi^{s^j_2} & q^{-D^j_2}  \\ \end{bmatrix}  \bigg\} 
 \text{, } 
\end{align*} }

\noindent the entries for the product of two L-operators introduced above take the form,

{\small \begin{align*}
  \underline{\mathcal{I}_1} \equiv          q^{D^j_1 + D^j_2} + \big( a^j_1 \big)^{\dagger} q^{D^j_1 } \xi^{s^j_1} q^{-2} a^j_2 q^{-D^j_2 - D^j_3 } \xi^{s-s^j_2}        \text{, } \end{align*}

  \begin{align*} \underline{\mathcal{I}_2} \equiv    q^{D^j_1} \big( a^j_2 \big)^{\dagger} q^{D^j_2} \xi^{s^j_2} + \big( a^j_1 \big)^{\dagger}   q^{D^j_1} \xi^{s^j_1} q^{-D^j_2 + D^j_3}   
 - \big( a^j_1 \big)^{\dagger} q^{D^j_1 } \xi^{s^j_1} q^{-2} q^{D^j_2 - 3 d^j_3 } \xi^s        \text{, }   \end{align*}

    \begin{align*} \underline{\mathcal{I}_3} \equiv        \big( a^j_1 \big)^{\dagger} q^{D^j_1} \xi^{s^j_1} \big( a^j_3 \big)^{\dagger} q^{D^j_3 } \xi^{s^j_3 }   \text{, }   \end{align*}}

\begin{figure}[H]
\begin{align*}
\includegraphics[width=0.7\columnwidth]{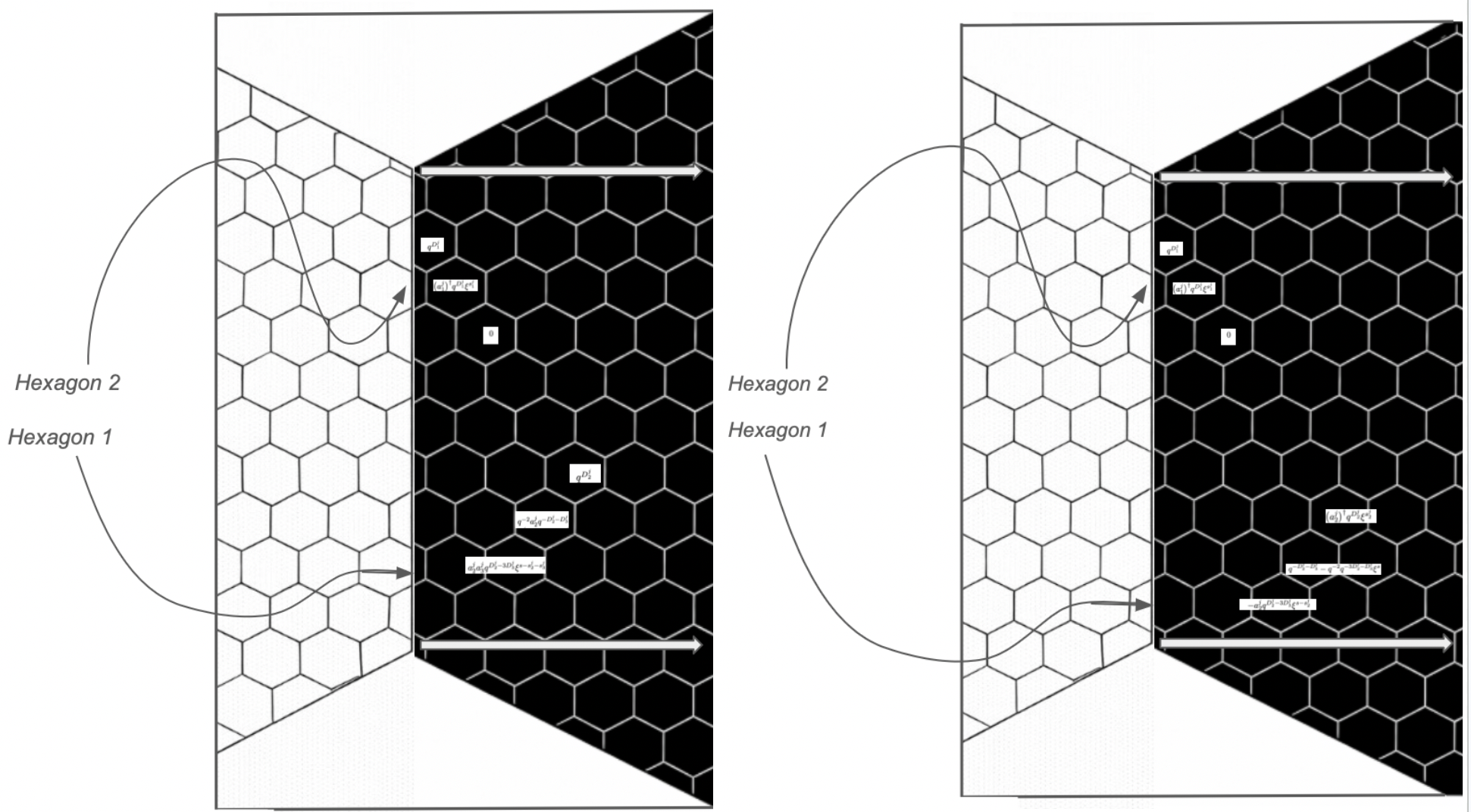}
\end{align*}
\caption{A depiction of the weak finite volume limit procedure associated with the product $\mathcal{T}_1 \mathcal{T}_2$ of two L-operators for the 20-vertex model.}
\end{figure}

 {\tiny  \begin{align*} \left\{\!\begin{array}{ll@{}>{{}}l}  \underline{\mathcal{I}_4} \equiv \mathcal{I}^1_4 + \mathcal{I}^2_4  , \\ \\ \mathcal{I}^1_4 \equiv q^{-2} a^j_1 q^{-D^j_1 - D^j_2} \xi^{s-s^j_1} q^{D^j_2} ,\\ \\ \mathcal{I}^2_4 \equiv q^{-D^j_1 + D^j_2} q^{-2} a^j_2 q^{-D^j_2 - D^j_3 } \xi^{s-s^j_2} , \\ \\ \mathcal{I}^3_4 \equiv - q^{-2} q^{D^j_1 - 3 D^j_2 } \xi^s q^{-2} a^j_2 q^{-D^j_2 - D^j_3} \xi^{s-s^j_2} , \\ \\ \mathcal{I}^4_4 \equiv \big( a^j_2 \big)^{\dagger}  q^{D^j_2} \xi^{s^j_2} a^j_2 a^j_3  q^{-D^j_2 - 3 D^j_3} \xi^{s-s^j_2 - s^j_3}  \text{ , }  \end{array}\right. \end{align*}}

 {\tiny \begin{align*}  \left\{\!\begin{array}{ll@{}>{{}}l}  \underline{\mathcal{I}_5} \equiv     \mathcal{I}^1_5 + \mathcal{I}^2_5 + \mathcal{I}^3_5  ,  \\ \\   \mathcal{I}^1_5 \equiv   q^{-2} a^j_1 q^{-D^j_1 - D^j_2} \xi^{s-s^j_1} \big( a^j_2 \big)^{\dagger} q^{D^j_2} \xi^{s^j_2} ,    \\ \\  \mathcal{I}^2_5 \equiv    q^{-D^j_1 + D^j_2 } q^{-D^j_2 + D^j_3} - q^{-2} q^{D^j_1 - 3 D^j_2 } \xi^s  q^{-D^j_2 + D^j_3} , \\ \\ \mathcal{I}^3_5 \equiv   - q^{-D^j_1 + D^j_2} q^{-2} q^{D^j_2 - 3 D^j_3}   \xi^s + q^{-2} q^{D^j_1 - 3 D^j_2} \xi^s q^{-2} q^{D^j_2 - 3 D^j_3} \xi^s ,   \\ \\  \mathcal{I}^4_5 \equiv  \big( a^j_2 \big)^{\dagger} q^{D^j_2} \xi^{s^j_2} \big( - a^j_3 \big) q^{D^j_3 - 3 D^j_4} \xi^{s-s^j_3}    \text{, }  \end{array}\right. \end{align*}

  \begin{align*} 
  \left\{\!\begin{array}{ll@{}>{{}}l}  \underline{\mathcal{I}_6} \equiv \mathcal{I}^1_6 + \mathcal{I}^2_6 \\ \\   \mathcal{I}^1_6 \equiv    q^{-D^j_1 + D^j_2} \big( a^j_3 \big)^{\dagger} q^{D^j_3} \xi^{s^j_3} - q^{-2} q^{D^j_1 - 3 D^j_2} \xi^s  \big( a^j_3 \big)^{\dagger} q^{D^j_3} \xi^{s^j_3} , \\ \\ \mathcal{I}^2_6 \equiv      \big( a^j_2 \big)^{\dagger}   q^{D^j_2} \xi^{s^j_2} q^{-D^j_3}         \text{, }  \end{array}\right. \end{align*}

  {\tiny \begin{align*} \left\{\!\begin{array}{ll@{}>{{}}l} \underline{\mathcal{I}_7} \equiv  \mathcal{I}^1_7 + \mathcal{I}^2_7  + \mathcal{I}^3_7 \\ \\  \mathcal{I}^1_7 \equiv  a^j_1 a^j_2 q^{-D^j_1 - 3 D^j_2} \xi^{s-s^j_1 - s^j_2} q^{D^j_2} \\ \\  \mathcal{I}^2_7 \equiv - a^j_2 q^{D^j_2 - 3 D^j_3} \xi^{s-s^j_2} q^{-2} a^j_2 q^{-D^j_2 - D^j_3} \xi^{s-s^j_2}  \\ \\ \mathcal{I}^3_7 \equiv q^{-D^j_2} a^j_2 a^j_3 q^{-D^j_2 - 3 D^j_2} \xi^{s-s^j_2 - s^j_3}    \text{, }  \end{array}\right. \end{align*} }

  \begin{align*}    \left\{\!\begin{array}{ll@{}>{{}}l} \underline{\mathcal{I}_8} \equiv  \mathcal{I}^1_8 + \mathcal{I}^2_8 + \mathcal{I}^3_8 ,   \\ \\   \mathcal{I}^1_8 \equiv  a^j_1 a^j_2 q^{-D^j_1 - 3 D^j_2} \xi^{s-s^j_1 - s^j_2} \big( a^j_2 \big)^{\dagger} q^{D^j_2} \xi^{s^j_2} , \\ \\ \mathcal{I}^2_8 \equiv  - a^j_2 q^{D^j_2 - 3 D^j_3} \xi^{s-s^j_2} \bigg[  q^{-D^j_2 + D^j_3} -  q^{-2 } q^{D^j_2 - 3 D^j_3} \xi^s \bigg] , \\ \\ \mathcal{I}^3_8 \equiv   - q^{-D^j_2}  a^j_3   q^{D^j_3 - 3 D^j_4}  \xi^{s-s^j_3}    \text{, }  \end{array}\right. 
  \end{align*}

 \begin{align*} \underline{\mathcal{I}_9}  \equiv         - a^j_2 q^{D^j_2 - 3 D^j_3} \xi^{s-s^j_2} \big( a^j_3 \big)^{\dagger} q^{D^j_3} \xi^{s^j_3} + q^{-D^j_2 - D^j_3}     \text{, }    
\end{align*} }

\noindent which are entries of,

\[
\begin{bmatrix}
\textbf{1}_2 & \textbf{4}_2 & \textbf{7}_2 \\ \textbf{2}_2 & \textbf{5}_2 & \textbf{8}_2 \\ \textbf{3}_2 & \textbf{6}_2 & \textbf{9}_2 
\end{bmatrix} \text{, }
\]

\noindent associated with the infinite volume weak limit as $\underline{M} \longrightarrow + \infty$, $N \longrightarrow + \infty$,

\[\underset{-N\leq i \leq 1}{\prod}\begin{bmatrix}
  \textbf{1}^i & \textbf{4}^i & \textbf{7}^i     \\   \textbf{2}^i & \textbf{5}^i & \textbf{8}^i   \\ \textbf{3}^i & \textbf{6}^i & \textbf{9}^i 
    \end{bmatrix} \text{ , }   \]

\noindent from which we conclude the argument. \boxed{}

\bigskip

\noindent \textbf{Lemma} \textit{2} (\textit{the product representation when varying two spectral parameters of the twenty-vertex model, a second base case induction step for lower order terms of the three-dimensional expansion}). Suppose that \textbf{Lemma} \textit{20V-CP-1} and \textbf{Lemma} \textit{20V-CP-2} hold. One has that the product of,

{\tiny \[
\begin{bmatrix}
     q^{D^1_1} & q^{-2} a^1_1 q^{-D^1_1 - D^1_2} \xi^{s-s^1_1} & a^1_1 a^1_2 q^{-D^1_1 - 3 D^1_2} \xi^{s-s^1_2 - s^2_2} \\ \big( a^1_1 \big)^{\dagger} q^{D^1_1} \xi^{s^1_1} & q^{-D^1_1 + D^1_2}  - q^{-2} q^{D^1_1 - 3 D^1_2} \xi^s  & - a^1_2 q^{D^1_2 - 3 D^1_3} \xi^{s-s^1_2} \\ 0 & \big( a^1_2 \big)^{\dagger} q^{D^1_2 } \xi^{s^1_2} & q^{-D^1_2}      \\
\end{bmatrix}  \text{, }
\]}

\noindent with,

{\tiny \[
 \begin{bmatrix}
  q^{D^2_1} & q^{-2} a^2_1 q^{-D^2_1 - D^2_2} \xi^{s-s^2_1} & a^2_1 a^2_2 q^{-D^2_1 - 3 D^2_2} \xi^{s-s^2_2 - s^3_2} \\ \big( a^2_1 \big)^{\dagger} q^{D^2_1} \xi^{s^2_1} & q^{-D^2_1 + D^2_2} - q^{-2} q^{D^2_1 - 3 D^2_2} \xi^s & - a^2_2 q^{D^2_2 - 3 D^2_3} \xi^{s-s^2_2} \\ 0 & \big( a^2_2 \big)^{\dagger} q^{D^2_2} \xi^{s^2_2} & q^{-D^2_2}   \\
\end{bmatrix}  \text{, }
\]}

\noindent can be expressed with the union of the span of the three subspaces,

 {\tiny  \[ \begin{bmatrix}
 q^{D^1_1 + D^2_1} \big( a^1_1 \big)^{\dagger} q^{D^1_1} \xi^{s^1_1} q^{-2} a^2_1 q^{-D^2_1 - D^2_2} \xi^{s-s^2_1} \\ \\ q^{D^1_1} \big( a^2_1 \big)^{\dagger} q^{D^2_1} \xi^s
 + \big( a^1_1 \big)^{\dagger} q^{D^1_1} \xi^{s^1_1} q^{-D^2_1 + D^2_2} - q^{-2} q^{D^2_1 - 3 D^2_2} \xi^s \\ \\  \big( a^1_1 \big)^{\dagger} q^{D^1_1} \xi^{s^1_1} \big( a^2_2 \big)^{\dagger} q^{D^2_2} \xi^{s^2_2}       
 \end{bmatrix} \text{, } \]

\begin{align*} \begin{bmatrix}      q^{D^1_1} q^{-2} a^2_1
 q^{-D^2_1 - D^2_2} \xi^{s-s^2_1} + q^{-2} a^1_1 q^{-D^1_1 - D^1_2} \xi^{s-s^1_1}   q^{-D^2_1 + D^2_2}  - q^{-2} a^1_1 q^{-D^1_1 - D^1_2} \xi^{s-s^1_1} q^{-2} q^{D^2_1 - 3 D^2_2} \xi^s    + a^1_1 a^1_2 q^{-D^1_1 - 3 D^1_2}   \\ \times \xi^{s-s^1_2 - s^2_2} \big( a^2_2 \big)^{\dagger} q^{D^2_2} \xi^{s^3_2} \\  \\  \big( a^1_1 \big)^{\dagger} q^{D^1_1} \xi^{s^1_1} q^{-2} a^2_1 q^{-D^2_1 - 3 D^2_2} \xi^{s-s^2_1} +    q^{-D^1_1 + D^1_2} q^{-D^2_1 + D^2_2} - q^{-2} q^{D^1_1 - 3 D^1_2} \xi^s  q^{-D^2_1 + D^2_2} - q^{-D^1_1 + D^1_2} q^{-2} q^{D^2_1 - 3 D^2_2} \xi^s   + q^{-2}  \\ \times q^{D^1_1 - 3 D^1_2} \xi^s q^{-2} q^{D^2_1 - 3 D^2_2} \xi^s  - a^1_2 q^{D^1_2 - 3 D^1_3} \xi^{s-s^1_2} \big( a^2_2 \big)^{\dagger} q^{D^2_2} \xi^{s^2_2}   \\   \\ \big( a^1_2 \big)^{\dagger} q^{D^1_2 } \xi^{s^1_2}  q^{-D^2_1 + D^2_2 } -  \big( a^1_2 \big)^{\dagger} q^{D^1_2 } q^{-2} q^{D^2_1 - 3 D^2_2} \xi^s  - a^1_2 q^{D^1_2 - 3 D^1_3} \xi^{s-s^1_2} \big( a^2_2 \big)^{\dagger} q^{D^2_2} \xi^{s^2_2}         
 \end{bmatrix} \text{, } \end{align*}

 \begin{align*}
 \begin{bmatrix}         q^{D^1_1} a^2_1 a^2_2 q^{-D^2_1 - 3 D^2_2} \xi^{s-s^2_2 - s^3_2 } + q^{-2} a^1_1 q^{-D^1_! - D^1_2 } \xi^{s-s^1_1}  \big( - a^2_2 \big) q^{D^2_2 - 3 D^2_3} \xi^{s-s^2_2}  + a^1_1 a^1_2 q^{-D^1_1 - 3 D^1_2}   \\  \\    \big( a^1_1 \big)^{\dagger} q^{D^1_1} \xi^{s^1_1} a^2_1 a^2_2 q^{-D^2_1 - 3 D^2_2} \xi^{s-s^2_2 - s^3_2} +  q^{-D^1_1 + D^1_2} \big( -a^2_2 \big) q^{D^2_2 - 3 D^2_3} \xi^{s-s^2_2}  - q^{-2}  \big( - a^2_2 \big) q^{D^2_2 - 3 D^2_3} \xi^{s-s^2_2} - a^1_2 q^{D^1_2 - 3 D^1_3}  \\ \times \xi^{s-s^1_2} q^{-D^2_2 }  \\ \\   \big( a^1_2 \big)^{\dagger} q^{D^1_2 
 } \xi^{s^1_2} \big( - a^2_2 \big) q^{D^2_2 - 3 D^2_3} \xi^{s-s^2_2} - a^1_2 q^{D^1_2 - 3 D^1_3} \xi^{s-s^1_2} q^{-D^2_2} 
 \end{bmatrix} \text{. }
\end{align*} } 

\noindent \textit{Proof of Lemma 2}. Directly manipulate the expressions obtained in the previous result above, from which we conclude the argument. \boxed{}

\subsection{Reformulating the system of Poisson brackets with entries of the three-dimensional product representation}

\noindent From finite-dimensional representations for products of L-operators of the form obtained in the previous section, the previously obtained system of $81$ Poisson brackets is equivalent, term by term, to the system of Poisson brackets,

{\small \[
\left\{\!\begin{array}{ll@{}>{{}}l} \boxed{(1)}:      \big\{  A {\tiny \big( \underline{u} \big)   }     , A {\tiny \big( \underline{u^{\prime}} \big) }   \big\} = \bigg\{     {\tiny \big[   \underline{\mathcal{T}^1_{(1,1)}} + \underline{\mathcal{T}^2_{(1,1)}}  + \underline{\mathcal{T}^3_{(1,1)}}  + \underline{\mathcal{T}^4_{(1,1)}}  + \underline{\mathcal{T}^5_{(1,1)}} + \underline{\mathcal{T}^6_{(1,1)}}      \big] \cdot  \big[  \underline{A_1 \big( \underline{u} \big) } 
        + \underline{A_2 \big( \underline{u} \big) } } \\  {\tiny + \underline{A_3 \big( \underline{u} \big) }  + \underline{A_4  \big( \underline{u} \big) }  + \underline{A_5 \big( \underline{u} \big) }  \bigg]    ,      \big[   
 \big( \underline{\mathcal{T}^1_{(1,1)}}\big)^{\prime} + \big( \underline{\mathcal{T}^2_{(1,1)}} \big)^{\prime}  +  \big(  \underline{\mathcal{T}^3_{(1,1)}}\big)^{\prime}  + \big( \underline{\mathcal{T}^4_{(1,1)}}\big)^{\prime}  + \big( \underline{\mathcal{T}^5_{(1,1)}}\big)^{\prime} + \big( \underline{\mathcal{T}^6_{(1,1)}}   \big)^{\prime}    \big]}  \\ {\tiny \times  \big[  \underline{A_1 \big( \underline{u^{\prime}} \big) } 
        + \underline{A_2 \big( \underline{u^{\prime}} \big) }  + \underline{A_3 \big( \underline{u^{\prime}} \big) }   + \underline{A_4  \big( \underline{u^{\prime}} \big) }  + \underline{A_5 \big( \underline{u^{\prime}} \big) }  \big]  }   \bigg\}
\text{ , } \\ \vdots 
\\  \boxed{(81)}: \big\{ I {\tiny \big( \underline{u} \big)} , I {\tiny \big( \underline{u^{\prime}} \big)} \big\} = \bigg\{     {\tiny    \big[  \underline{\mathcal{T}^1_{(3,3)}} + \underline{\mathcal{T}^2_{(3,3)} }+ \underline{\mathcal{T}^3_{(3,3)}}  + \underline{\mathcal{T}^4_{(3,3)}        }     \big] \cdot  \big[   \underline{I_1 \big( \underline{u} \big)}  + \underline{I_2 \big( \underline{u} \big)}   + \underline{I_3 \big( \underline{u} \big)}    + \underline{I_4 \big( \underline{u} \big)}  
            \big] , }  \\  {\tiny    \big[  \big( \underline{\mathcal{T}^1_{(3,3)}} \big)^{\prime} + \big( \underline{\mathcal{T}^2_{(3,3)} } \big)^{\prime}  + \big( \underline{\mathcal{T}^3_{(3,3)}}  \big)^{\prime} + \big( \underline{\mathcal{T}^4_{(3,3)}} \big)^{\prime}            \big] \cdot  \big[   \underline{I_1 \big( \underline{u^{\prime}} \big)}  + \underline{I_2 \big( \underline{u^{\prime}} \big)}   + \underline{I_3 \big( \underline{u^{\prime}} \big)}    + \underline{I_4 \big( \underline{u^{\prime}} \big)}  
            \big]    }                 \bigg\} \text{. }
\end{array}\right.
\] }

\subsection{Poisson bracket approximations in asymptotic weak finite volume}

\noindent With approximations for each entry of the three-dimensional product representation, below we demonstrate, for each Poisson bracket within the system, how terms can be asymptotically approximated in the weak finite volume limit. Each Poisson bracket can be simplified through the following sequence of steps:

{\small \begin{itemize}
    \item[$\bullet$] \boxed{(1)} : one application of the \underline{bilinearity}, (BL), property, to separate terms from $\mathcal{T}$, $A$, $B$, $C$, $D$, $E$, $F$, $G$, $H$, or $I$, factors,

    \item[$\bullet$] \boxed{(2)} : one application of the \underline{Leibniz rule}, (LR), to simplify the number of terms appearing in the first argument of the Poisson bracket from two to one,

    \item[$\bullet$] \boxed{(3)} : one application of the \underline{anticommutativity}, (AC), property, to reverse the order in which the arguments appear in the superposition of two Poisson brackets obtained after an application of (LR) in the previous step,

    \item[$\bullet$] \boxed{(4)} : one application of the \underline{Leibniz rule}, (LR), to simplify the number of arguments appearing in the first argument of the first Poisson bracket from the superposition of Poisson brackets obtained in the second step,

    \item[$\bullet$] \boxed{(5)} : a final application of the \underline{Leibniz rule}, (LR), to simplify the number of arguments appearing in the first argument of the second Poisson bracket from the superposition of Poisson brackets obtained in the second step,
    
\end{itemize}

\begin{itemize}
    \item[$\bullet$] \boxed{(6)} : taking a product over constants obtained for each Poisson bracket, to obtain each of the $81$ constants for approximating each Poisson bracket from the three-dimensional Poisson structure.
\end{itemize} }

\noindent After having performed each of the six steps above, we obtain the desired approximation for each Poisson bracket, within the three-dimensional family, by taking a product over approximations for Poisson brackets before the product of at most ${\tiny 4} \cdot  \lceil {\tiny \frac{m-4}{3} } \rceil $ terms was substituted in. For example, given the first bracket,

{\small  \begin{align*} \bigg\{     \big[ {\tiny   \underline{\mathcal{T}^1_{(1,1)}} + \underline{\mathcal{T}^2_{(1,1)}}  + \underline{\mathcal{T}^3_{(1,1)}}  + \underline{\mathcal{T}^4_{(1,1)}}  + \underline{\mathcal{T}^5_{(1,1)}} + \underline{\mathcal{T}^6_{(1,1)}}    }  \big]  \big[  {\tiny \underline{A_1 \big( \underline{u} \big) } 
        + \underline{A_2 \big( \underline{u} \big) }  + \underline{A_3 \big( \underline{u} \big) }  + \underline{A_4  \big( \underline{u} \big) }   + \underline{A_5 \big( \underline{u} \big) } }   \big] ,  \\          \big[ {\tiny   
 \big( \underline{\mathcal{T}^1_{(1,1)}}\big)^{\prime}  + \big( \underline{\mathcal{T}^2_{(1,1)}} \big)^{\prime}  +  \big(  \underline{\mathcal{T}^3_{(1,1)}}\big)^{\prime}  + \big( \underline{\mathcal{T}^4_{(1,1)}}\big)^{\prime}  + \big( \underline{\mathcal{T}^5_{(1,1)}}\big)^{\prime} + \big( \underline{\mathcal{T}^6_{(1,1)}}   \big)^{\prime}  }  \big]  \big[ {\tiny \underline{A_1 \big( \underline{u^{\prime}} \big) } 
        + \underline{A_2 \big( \underline{u^{\prime}} \big) }  + \underline{A_3 \big( \underline{u^{\prime}} \big) } } \\  {\tiny + \underline{A_4  \big( \underline{u^{\prime}} \big) }   + \underline{A_5 \big( \underline{u^{\prime}} \big) }  } \big]     \bigg\}
\text{, } \end{align*} }

\noindent can be used to asymptotically approximate,

 {\tiny  \begin{align*} \bigg\{     \big[  {\tiny \underline{\mathcal{T}^1_{(1,1)}} + \underline{\mathcal{T}^2_{(1,1)}}  + \underline{\mathcal{T}^3_{(1,1)}}  + \underline{\mathcal{T}^4_{(1,1)}}  + \underline{\mathcal{T}^5_{(1,1)}} + \underline{\mathcal{T}^6_{(1,1)}} }      \big] \cdot  \big[ {\tiny    \underset{i \in \textbf{N} :  |  i    | \leq  4 \lceil \frac{m-4}{3} \rceil }{\prod}   \underline{A_{1,i} \big( \underline{u}\big) }    +  \underset{i \in \textbf{N} :  |  i    | \leq  4 \lceil \frac{m-4}{3} \rceil }{\prod}   \underline{A_{2,i} \big( \underline{u}\big) }   }        \end{align*}

  \begin{align*} {\tiny \underset{i \in \textbf{N} :  |  i    | \leq  4 \lceil \frac{m-4}{3} \rceil }{\prod}   \underline{A_{3,i} \big( \underline{u}\big) }   + \underset{i \in \textbf{N} :  |  i    | \leq  4 \lceil \frac{m-4}{3} \rceil }{\prod}   \underline{A_{4,i} \big( \underline{u}\big) }  + \underset{i \in \textbf{N} :  |  i    | \leq  4 \lceil \frac{m-4}{3} \rceil }{\prod}   \underline{A_{5,i} \big( \underline{u}\big) } } \big]     ,      \big[  {\tiny 
 \big( \underline{\mathcal{T}^1_{(1,1)}}\big)^{\prime} + \big( \underline{\mathcal{T}^2_{(1,1)}} \big)^{\prime}  +  \big(  \underline{\mathcal{T}^3_{(1,1)}}\big)^{\prime}  } \end{align*}

  \begin{align*}  {\tiny + \big( \underline{\mathcal{T}^4_{(1,1)}}\big)^{\prime}  + \big( \underline{\mathcal{T}^5_{(1,1)}}\big)^{\prime} + \big( \underline{\mathcal{T}^6_{(1,1)}}   \big)^{\prime}    \big] \cdot  \big[     \underset{i \in \textbf{N} :  |  i    | \leq  4 \lceil \frac{m-4}{3} \rceil }{\prod}   \underline{A_{1,i} \big( \underline{u}\big) }   + \underset{i \in \textbf{N} :  |  i    | \leq  4 \lceil \frac{m-4}{3} \rceil }{\prod}   \underline{A_{2,i} \big( \underline{u}\big) }          + \underset{i \in \textbf{N} :  |  i    | \leq  4 \lceil \frac{m-4}{3} \rceil }{\prod}   \underline{A_{3,i} \big( \underline{u}\big) } }  \end{align*}

  \begin{align*}  {\tiny + \underset{i \in \textbf{N} :  |  i    | \leq  4 \lceil \frac{m-4}{3} \rceil }{\prod}   \underline{A_{4,i} \big( \underline{u}\big) }  + \underset{i \in \textbf{N} :  |  i    | \leq  4 \lceil \frac{m-4}{3} \rceil }{\prod}   \underline{A_{5,i} \big( \underline{u}\big) }  } \big]     \bigg\}
\text{, } \end{align*} }

\noindent from the observation that the bracket above is equivalent to, after having followed by the steps described at the beginning of the section,

{\small   \begin{align*}
  \bigg\{     \bigg[ \underset{k \in \textbf{N}: 1 \leq k \leq 6}{\sum}  {\tiny \underline{\mathcal{T}^k_{(1,1)}} }    \bigg] \cdot  \bigg[ \underset{j \in \textbf{N}: 1 \leq j \leq 5}{\sum}    \bigg[   \underset{i \in \textbf{N} :  |  i    | \leq  4 \lceil \frac{m-4}{3} \rceil }{\prod}   {\tiny \underline{A_{j,i} \big( \underline{u}\big) } } \bigg]   \bigg]     ,      \bigg[  \underset{k^{\prime} \in \textbf{N}: 1 \leq k \leq 6}{\sum}   
 \big( {\tiny \underline{\mathcal{T}^{k^{\prime}}_{(1,1)}} } \big)^{\prime}    \bigg] \\ \times  \bigg[ \underset{j^{\prime} \in \textbf{N}: 1 \leq j^{\prime} \leq 5}{\sum}    \bigg[    \underset{i \in \textbf{N} :  |  i    | \leq  4 \lceil \frac{m-4}{3} \rceil }{\prod}   {\tiny \underline{A_{j^{\prime},i} \big( \underline{u}\big) } } \bigg] \bigg]     \bigg\}  \end{align*}}

{\small  \begin{align*} \\ \overset{(\mathrm{BL})}{=} \underset{j^{\prime} \in \textbf{N} : 1 \leq j^{\prime} \leq 5}{\underset{k^{\prime} \in \textbf{N}: 1 \leq k \leq 6}{\underset{j \in \textbf{N}: 1 \leq j \leq 5}{\underset{k \in \textbf{N}: 1 \leq k \leq 6}{\sum}}}}  \bigg\{     {\tiny \underline{\mathcal{T}^k_{(1,1)}}   }    \cdot \bigg[   \underset{i \in \textbf{N} :  |  i    | \leq  4 \lceil \frac{m-4}{3} \rceil }{\prod}   {\tiny \underline{A_{j,i} \big( \underline{u}\big) } }  \bigg]        ,      
 \big( {\tiny \underline{\mathcal{T}^{k^{\prime}}_{(1,1)}}} \big)^{\prime}   \cdot   \bigg[    \underset{i \in \textbf{N} :  |  i    | \leq  4 \lceil \frac{m-4}{3} \rceil }{\prod}   {\tiny \underline{A_{j^{\prime},i} \big( \underline{u}\big) } } \bigg]     \bigg\} \\  \vdots \end{align*} }   
  
  {\tiny \begin{align*} \overset{(\mathrm{LR})}{=} -          \underset{j^{\prime} \in \textbf{N} : 1 \leq j^{\prime} \leq 5}{\underset{k^{\prime} \in \textbf{N}: 1 \leq k \leq 6}{\underset{j \in \textbf{N}: 1 \leq j \leq 5}{\underset{k \in \textbf{N}: 1 \leq k \leq 6}{\sum}}}}                  \bigg[  \bigg\{            \big( \mathcal{T}^{i^{\prime\prime}}_{(1,1)}  \big)^{\prime}                     , \underline{\mathcal{T}^i_{(1,1)}}           \bigg\} \bigg[  {\underset{j \neq j^{\prime} \in \textbf{N} :  |  i    | , |j |  , |j^{\prime} |  \leq  4 \lceil \frac{m-4}{3} \rceil }{\prod}}                             \underline{A_{j,i} \big( \underline{u}\big) } \text{ }   \underline{A_{j^{\prime},i} \big( \underline{u}\big) }       \bigg]   +   \bigg( \big( \mathcal{T}^{i^{\prime\prime}}_{(1,1)}  \big)^{\prime}   \bigg)      \bigg\{                                         \underset{j \neq j^{\prime} \in \textbf{N}: |j | , | j^{\prime}| \leq 4 \lceil \frac{m-4}{3} \rceil }{\underset{i \in \textbf{N} :  |  i    | \leq  4 \lceil \frac{m-4}{3} \rceil }{\prod}}                             \underline{A_{j,i} \big( \underline{u}\big) }    \\   , \underline{\mathcal{T}^i_{(1,1)}}           \bigg\}    \bigg[   {\underset{i,j \in \textbf{N} :  |  i    | \leq  4 \lceil \frac{m-4}{3} \rceil }{\prod}}                             \underline{A_{j,i} \big( \underline{u}\big) }  \bigg]  +   \bigg(  \underline{\mathcal{T}^i_{(1,1)}}  \bigg)    \bigg[   \bigg\{   \big( \mathcal{T}^{i^{\prime\prime}}_{(1,1)}  \big)^{\prime}           ,           \underset{j \in \textbf{N} :  |  i    | \leq  4 \lceil \frac{m-4}{3} \rceil }{\prod}              \underline{A_{j,i} \big( \underline{u}\big) }            \bigg\}  \\ \times   \bigg[     \underset{j^{\prime} \in \textbf{N} :  |  i    | \leq  4 \lceil \frac{m-4}{3} \rceil }{\prod}              \underline{A_{j^{\prime},i} \big( \underline{u}\big) }         \bigg]   
   +    \bigg( \big( \mathcal{T}^{i^{\prime\prime}}_{(1,1)}  \big)^{\prime}   \bigg)  
{\underset{i \neq j \neq j^{\prime} \in \textbf{N} :  |  i    | , |j |  , |j^{\prime} |  \leq  4 \lceil \frac{m-4}{3} \rceil }{\prod}}  \bigg\{                            \underline{A_{j,i} \big( \underline{u}\big) } ,       \underline{A_{j^{\prime},i} \big( \underline{u}\big) }              \bigg\}  \bigg]        \bigg]     \text{. }           
\end{align*} }

\noindent The asymptotic approximation for Poisson brackets of the form above, namely brackets involving products of $A$, $B$, $C$, $D$, $E$, $F$, $G$, $H$, or $I$, terms, above is obtained from manipulating a simpler Poisson bracket before the product of terms over $i,j,j^{\prime}, k$ or $k^{\prime}$ is inserted into expressions, from the previous section, that were introduced for each of the 9 entries of the L-operator product representation.

\subsection{Argument of the main result (3D Poisson structure) from block operators of the transfer matrix for the $20$-vertex model}

\noindent \textit{Proof of Lemma P1}. Directly apply the arguments for $\textbf{Lemma}$ $\textit{1}$ and $\textbf{Lemma}$ $\textit{2}$ in the previous section, from which we conclude the argument. \boxed{}

\bigskip 

\noindent \textit{Proof of Main Result}. 
\noindent We make use of the finite-dimensional product representation for the transfer matrix obtained in $\textbf{Lemma}$ $\textit{P1}$, $\textit{P2}$ and $\textit{P3}$. We perform the following computations by making use of the above list of properties from the Poisson bracket. For simplicity, we denote the constants used for each bracket with $C_1$, $C_2$, and $C_{1,2,3}$, which can be trivially relabeled:

        \subsubsection{Poisson brackets of Type $1$ away from the diagonal}

        {\small \begin{align*}
        \underline{\textit{Poisson bracket of Type 1}:}  \bigg\{   \big[ {\tiny   \underline{\mathcal{T}^1_{(1,1)}} + \underline{\mathcal{T}^2_{(1,1)}}  + \underline{\mathcal{T}^3_{(1,1)}}  + \underline{\mathcal{T}^4_{(1,1)}}  + \underline{\mathcal{T}^5_{(1,1)}} + \underline{\mathcal{T}^6_{(1,1)}}    }  \big] \cdot   \big[  {\tiny \underline{A_1 \big( \underline{u} \big) } 
        + \underline{A_2 \big( \underline{u} \big) }}  \\  + {\tiny \underline{A_3 \big( \underline{u} \big) }  + \underline{A_4  \big( \underline{u} \big) }   + \underline{A_5 \big( \underline{u} \big) } } \big]   ,   \big[              {\tiny    \big( \underline{\mathcal{T}^1_{(1,2)} } \big)^{\prime}   +             \big(  \underline{\mathcal{T}^2_{(1,2)}} \big)^{\prime}   + \big( \underline{\mathcal{T}^3_{(1,2)}}  \big)^{\prime} +  \big( \underline{\mathcal{T}^4_{(1,2)}}  \big)^{\prime} }  \big] \cdot  \big[   {\tiny \underline{B_1 \big( \underline{u^{\prime}} \big) }          + \underline{B_2 \big( \underline{u^{\prime}} \big) } } \\  {\tiny + \underline{B_3 \big( \underline{u^{\prime}} \big) }    + \underline{B_4 \big( \underline{u^{\prime}} \big) }   } \big]            \bigg\}   \end{align*}

   \begin{align*} = \big\{        \big( \underset{1 \leq i \leq 6}{\sum} {\tiny \underline{\mathcal{T}^i_{(1,1)}}  \big)} \cdot  \big( \underset{1 \leq i^{\prime}\leq 5}{\sum}   {\tiny \underline{A_{i^{\prime}} \big( \underline{u} \big)}  }        \big)      ,     \big( \underset{1 \leq i^{\prime\prime} \leq 4}{\sum}     {\tiny \big( \mathcal{T}^{i^{\prime\prime}}_{(1,2)}  \big)^{\prime}   \big) } \cdot  \big(      \underset{1 \leq i^{\prime\prime\prime} \leq 4}{\sum}                        {\tiny \underline{B_{i^{\prime\prime\prime}} \big( \underline{u} \big) } }       \big)            \big\}  \\   \end{align*}

   \begin{align*}  \overset{(\mathrm{BL})}{=}    \underset{i,i^{\prime},i^{\prime\prime},i^{\prime\prime\prime}}{\sum}  \big\{         {\tiny \big(  \underline{\mathcal{T}^i_{(1,1)}}  }\big) \cdot \big( {\tiny \underline{A_{i^{\prime}} \big( \underline{u} \big)}  }       \big)    , \big( {\tiny \big( \mathcal{T}^{i^{\prime\prime}}_{(1,2)}  \big)^{\prime} }  \big) \cdot   \big( {\tiny \underline{B_{i^{\prime\prime\prime}} \big( \underline{u} \big) }  }     \big) \big\} \\ \\   \overset{(\mathrm{LR})}{=}    \underset{i,i^{\prime},i^{\prime\prime},i^{\prime\prime\prime}}{\sum}  
 \bigg[ {\tiny \big\{         \underline{\mathcal{T}^i_{(1,1)}} , \big( \big( \mathcal{T}^{i^{\prime\prime}}_{(1,2)}  \big)^{\prime}   \big)  \big( \underline{B_{i^{\prime\prime\prime}} \big( \underline{u} \big) }       \big) \big\}  \big( \underline{A_{i^{\prime}} \big( \underline{u} \big)}         \big)    + \big(  \underline{\mathcal{T}^i_{(1,1)}}  \big)  \big\{         \underline{A_{i^{\prime}} \big( \underline{u} \big)}           , \big( \big( \mathcal{T}^{i^{\prime\prime}}_{(1,2)}  \big)^{\prime}   \big)  \big( \underline{B_{i^{\prime\prime\prime}} \big( \underline{u} \big) }       \big) \big\}  }  \bigg] \\ \\    \overset{(\mathrm{AC})}{=}   \underset{i,i^{\prime},i^{\prime\prime},i^{\prime\prime\prime}}{\sum}  
 \bigg[ {\tiny - \big\{          \big( \big( \mathcal{T}^{i^{\prime\prime}}_{(1,2)}  \big)^{\prime}   \big)  \big( \underline{B_{i^{\prime\prime\prime}} \big( \underline{u} \big) }       \big) , \underline{\mathcal{T}^i_{(1,1)}} \big\}  \big( \underline{A_{i^{\prime}} \big( \underline{u} \big)}         \big)    - \big(  \underline{\mathcal{T}^i_{(1,1)}}  \big)  \big\{     \big( \big( \mathcal{T}^{i^{\prime\prime}}_{(1,2)}  \big)^{\prime}   \big)  \big( \underline{B_{i^{\prime\prime\prime}} \big( \underline{u} \big) }       \big)  ,     \underline{A_{i^{\prime}} \big( \underline{u} \big)}           \big\} }   \bigg]  \end{align*}

 \begin{align*}
 \overset{(\mathrm{LR})}{=}  -  \underset{i,i^{\prime},i^{\prime\prime},i^{\prime\prime\prime}}{\sum}  
 \bigg[  {\tiny \big\{           \big( \mathcal{T}^{i^{\prime\prime}}_{(1,2)}  \big)^{\prime}     , \underline{\mathcal{T}^i_{(1,1)}} \big\} \big( \underline{B_{i^{\prime\prime\prime}} \big( \underline{u} \big) }       \big)  \big( \underline{A_{i^{\prime}} \big( \underline{u} \big)}         \big)  +    \big( \big( \mathcal{T}^{i^{\prime\prime}}_{(1,2)}  \big)^{\prime}   \big) \big\{        \underline{B_{i^{\prime\prime\prime}} \big( \underline{u} \big) }      , \underline{\mathcal{T}^i_{(1,1)}} \big\}  \big( \underline{A_{i^{\prime}} \big( \underline{u} \big)}         \big) } \\ {\tiny +  \big(  \underline{\mathcal{T}^i_{(1,1)}}  \big)  \big\{     \big( \big( \mathcal{T}^{i^{\prime\prime}}_{(1,2)}  \big)^{\prime}   \big)  \big( \underline{B_{i^{\prime\prime\prime}} \big( \underline{u} \big) }       \big)  ,     \underline{A_{i^{\prime}} \big( \underline{u} \big)}           \big\}  }  \bigg] \end{align*}

 \begin{align*}       \overset{(\mathrm{LR})}{=}  -  \underset{i,i^{\prime},i^{\prime\prime},i^{\prime\prime\prime}}{\sum}   
 \bigg[ {\tiny \big\{           \big( \mathcal{T}^{i^{\prime\prime}}_{(1,2)}  \big)^{\prime}     , \underline{\mathcal{T}^i_{(1,1)}} \big\} \big( \underline{B_{i^{\prime\prime\prime}} \big( \underline{u} \big) }       \big)  \big( \underline{A_{i^{\prime}} \big( \underline{u} \big)}         \big)  +    \big( \big( \mathcal{T}^{i^{\prime\prime}}_{(1,2)}  \big)^{\prime}   \big) \big\{        \underline{B_{i^{\prime\prime\prime}} \big( \underline{u} \big) }      , \underline{\mathcal{T}^i_{(1,1)}} \big\}  \big( \underline{A_{i^{\prime}} \big( \underline{u} \big)}         \big) } \\ {\tiny +  \big(  \underline{\mathcal{T}^i_{(1,1)}}  \big) \bigg[  \big\{    \big( \mathcal{T}^{i^{\prime\prime}}_{(1,2)}  \big)^{\prime}      ,     \underline{A_{i^{\prime}} \big( \underline{u} \big)}           \big\} \big( \underline{B_{i^{\prime\prime\prime}} \big( \underline{u} \big) }       \big)    +  \big( \big( \mathcal{T}^{i^{\prime\prime}}_{(1,2)}  \big)^{\prime}   \big) \big\{      \underline{B_{i^{\prime\prime\prime}} \big( \underline{u} \big) }       ,     \underline{A_{i^{\prime}} \big( \underline{u} \big)}           \big\} \bigg]  }  \bigg]  \text{. } \end{align*} }

  \noindent For the first two terms of the superposition superposition above, $(*)$, there exists two strictly negative constants, $C_1$ and $C_2$, together implying that the first two terms can be approximated with,

        {\small \begin{align*}
       (*)  \approx  -\underset{i,i^{\prime}, i^{\prime\prime} , i^{\prime\prime\prime}}{\sum}  \bigg[ \frac{                     \underline{B_{i^{\prime\prime\prime}} \big( \underline{u} \big) }  \text{ }  \underline{A_{i^{\prime}} \big( \underline{u} \big)}       }{i^{\prime\prime}_{(1,2)} - i_{(1,2)} }  +       \big( \mathcal{T}^{i^{\prime\prime}}_{(1,1)}  \big)^{\prime} \bigg]  \equiv  - \underset{i,i^{\prime}, i^{\prime\prime} , i^{\prime\prime\prime}}{\sum}  
 \bigg[ \frac{                     \underline{B_{i^{\prime\prime\prime}} \big( \underline{u} \big) }  \text{ }  \underline{A_{i^{\prime}} \big( \underline{u} \big)}       }{i^{\prime\prime} - i }  +     \big( \mathcal{T}^{i^{\prime\prime}}_{(1,1)}  \big)^{\prime}  \bigg]  \approx \frac{C_1}{i^{\prime\prime}-i} + C_2         \text{, }  \end{align*}  }

        \noindent corresponding to the first, and last, bracket appearing in $(*)$, and another strictly negative constant,

         {\small \begin{align*}
                  C_{1,2,3}      \text{, }
            \end{align*} }

        \noindent corresponding to the remaining two brackets, from the observation,

            {\small \begin{align*}
              \big( \big( \mathcal{T}^{i^{\prime\prime}}_{(1,2)}  \big)^{\prime}   \big) \big\{        \underline{B_{i^{\prime\prime\prime}} \big( \underline{u} \big) }      , \underline{\mathcal{T}^i_{(1,1)}} \big\}  \big( \underline{A_{i^{\prime}} \big( \underline{u} \big)}         \big)   +     \big(  \underline{\mathcal{T}^i_{(1,1)}}  \big)   \big\{    \big( \mathcal{T}^{i^{\prime\prime}}_{(1,2)}  \big)^{\prime}      ,     \underline{A_{i^{\prime}} \big( \underline{u} \big)}           \big\} \big( \underline{B_{i^{\prime\prime\prime}} \big( \underline{u} \big) }       \big)         \approx \big[ \big(   \big( \mathcal{T}^{i^{\prime\prime}}_{(1,2)}  \big)^{\prime}   +  \underline{\mathcal{T}^i_{(1,1)}}   \big)  \big]      \\  \times \big[  \big\{        \underline{B_{i^{\prime\prime\prime}} \big( \underline{u} \big) }      , \underline{\mathcal{T}^i_{(1,1)}} \big\}     +  \big\{    \big( \mathcal{T}^{i^{\prime\prime}}_{(1,2)}  \big)^{\prime}      ,     \underline{A_{i^{\prime}} \big( \underline{u} \big)}           \big\} \big] \big[ \big(  \underline{A_{i^{\prime}} \big( \underline{u} \big)}      +  \underline{B_{i^{\prime\prime\prime}} \big( \underline{u} \big) }    \big)  \big]   \approx C_{1,2,3}    \text{. }
            \end{align*} }

    \subsubsection{Poisson brackets of Type $2$ on the diagonal}
            
            {\small \begin{align*}
          \underline{\textit{Poisson bracket of Type 2}:}  \bigg\{         \big[ {\tiny  \underline{\mathcal{T}^1_{(3,3)}} + \underline{\mathcal{T}^2_{(3,3)} }+ \underline{\mathcal{T}^3_{(3,3)}}  + \underline{\mathcal{T}^4_{(3,3)}        }  }   \big] \cdot  \big[   {\tiny \underline{I_1 \big( \underline{u} \big)}  + \underline{I_2 \big( \underline{u} \big)}   + \underline{I_3 \big( \underline{u} \big)}    + \underline{I_4 \big( \underline{u} \big)}  
         }   \big]  \\  ,   \big[  {\tiny \big( \underline{\mathcal{T}^1_{(3,3)}} \big)^{\prime} + \big( \underline{\mathcal{T}^2_{(3,3)} } \big)^{\prime}  + \big( \underline{\mathcal{T}^3_{(3,3)}}  \big)^{\prime} + \big( \underline{\mathcal{T}^4_{(3,3)}} \big)^{\prime}      }      \big] \cdot  \big[  {\tiny \underline{I_1 \big( \underline{u^{\prime}} \big)}  + \underline{I_2 \big( \underline{u^{\prime}} \big)}   + \underline{I_3 \big( \underline{u^{\prime}} \big)}    + \underline{I_4 \big( \underline{u^{\prime}} \big)}  
          }   \big]                    \bigg\} \end{align*}

            \begin{align*} 
            {=}  \big\{   \big( \underset{1 \leq i \leq 4}{\sum} {\tiny \underline{\mathcal{T}^i_{(3,3)}} } \big) \cdot  \big( \underset{1 \leq i^{\prime}\leq 4}{\sum}  {\tiny \underline{I_{i^{\prime}} \big( \underline{u} \big)}    }      \big)   ,    \big(    \underset{1 \leq i^{\prime\prime} \leq 4}{\sum}     {\tiny \underline{\mathcal{T}^{i^{\prime\prime}}_{(3,3)}}               }       \big) \cdot  \big(    \underset{1 \leq i^{\prime\prime\prime} \leq 4}{\sum}  {\tiny     \underline{I_{i^{\prime\prime\prime}} \big( \underline{u} \big) }  }       \big)    \big\} \\ \\ 
            \overset{(\mathrm{BL})}{=}    \underset{i,i^{\prime},i^{\prime\prime},i^{\prime\prime\prime}}{\sum}  \big\{    {\tiny  \big(        \underline{\mathcal{T}^i_{(3,3)}}       \big) \cdot \big(  \underline{I_{i^{\prime}} \big( \underline{u} \big)}            \big)               ,   \big( \underline{\mathcal{T}^{i^{\prime\prime}}_{(3,3)}}                     \big)   \cdot                         \big(  \underline{I_{i^{\prime\prime\prime}} \big( \underline{u} \big) }        \big)  }   \big\} \end{align*}
            
            \begin{align*} \overset{(\mathrm{LR})}{=}   \underset{i,i^{\prime},i^{\prime\prime},i^{\prime\prime\prime}}{\sum}  \bigg[ {\tiny \big\{    \underline{\mathcal{T}^i_{(3,3)}}  ,       \big( \underline{\mathcal{T}^{i^{\prime\prime}}_{(3,3)}}                     \big)                           \big(  \underline{I_{i^{\prime\prime\prime}} \big( \underline{u} \big) }        \big)          \big\}      \big(  \underline{I_{i^{\prime}} \big( \underline{u} \big)}            \big)            +     \big(   \underline{\mathcal{T}^i_{(3,3)}}  \big)   \big\{       \underline{I_{i^{\prime}} \big( \underline{u} \big)}          ,      \big( \underline{\mathcal{T}^{i^{\prime\prime}}_{(3,3)}}                     \big)                           \big(  \underline{I_{i^{\prime\prime\prime}} \big( \underline{u} \big) }        \big)         \big\}  }                  \bigg] \\ \\  \overset{(\mathrm{AC})}{=}              \underset{i,i^{\prime},i^{\prime\prime},i^{\prime\prime\prime}}{\sum} \bigg[ {\tiny - \big\{          \big( \underline{\mathcal{T}^{i^{\prime\prime}}_{(3,3)}}                     \big)                           \big(  \underline{I_{i^{\prime\prime\prime}} \big( \underline{u} \big) }        \big)    , \underline{\mathcal{T}^i_{(3,3)}}         \big\}      \big(  \underline{I_{i^{\prime}} \big( \underline{u} \big)}            \big)          -     \big(   \underline{\mathcal{T}^i_{(3,3)}}  \big)   \big\{                \big( \underline{\mathcal{T}^{i^{\prime\prime}}_{(3,3)}}                     \big)                           \big(  \underline{I_{i^{\prime\prime\prime}} \big( \underline{u} \big) }        \big)    ,   \underline{I_{i^{\prime}} \big( \underline{u} \big)}          \big\} }                   \bigg]   \\ \\   \overset{(\mathrm{LR})}{=}  -  \underset{i,i^{\prime},i^{\prime\prime},i^{\prime\prime\prime}}{\sum}        \bigg[      {\tiny \big\{      \underline{\mathcal{T}^{i^{\prime\prime}}_{(3,3)}}                                             , \underline{\mathcal{T}^i_{(3,3)}}         \big\}  \big(  \underline{I_{i^{\prime\prime\prime}} \big( \underline{u} \big) }        \big) \big(  \underline{I_{i^{\prime}} \big( \underline{u} \big)}            \big)   +                  \big( \underline{\mathcal{T}^{i^{\prime\prime}}_{(3,3)}}                     \big)      \big\{                                   \underline{I_{i^{\prime\prime\prime}} \big( \underline{u} \big) }          , \underline{\mathcal{T}^i_{(3,3)}}         \big\} \big(  \underline{I_{i^{\prime}} \big( \underline{u} \big)}            \big) } \\  {\tiny +      \big(   \underline{\mathcal{T}^i_{(3,3)}}  \big)   \big\{                \big( \underline{\mathcal{T}^{i^{\prime\prime}}_{(3,3)}}                     \big)                           \big(  \underline{I_{i^{\prime\prime\prime}} \big( \underline{u} \big) }        \big)    ,   \underline{I_{i^{\prime}} \big( \underline{u} \big)}          \big\} }                        \bigg]  \end{align*}

            \begin{align*}   \overset{(\mathrm{LR})}{=}      -  \underset{i,i^{\prime},i^{\prime\prime},i^{\prime\prime\prime}}{\sum}        \bigg[       {\tiny \big\{      \underline{\mathcal{T}^{i^{\prime\prime}}_{(3,3)}}                                             , \underline{\mathcal{T}^i_{(3,3)}}         \big\}  \big(  \underline{I_{i^{\prime\prime\prime}} \big( \underline{u} \big) }        \big) \big(  \underline{I_{i^{\prime}} \big( \underline{u} \big)}            \big)   +                  \big( \underline{\mathcal{T}^{i^{\prime\prime}}_{(3,3)}}                     \big)      \big\{                                   \underline{I_{i^{\prime\prime\prime}} \big( \underline{u} \big) }          , \underline{\mathcal{T}^i_{(3,3)}}         \big\} \big(  \underline{I_{i^{\prime}} \big( \underline{u} \big)}            \big)} \\ {\tiny +      \big(   \underline{\mathcal{T}^i_{(3,3)}}  \big)                       \bigg[       \big\{                \underline{\mathcal{T}^{i^{\prime\prime}}_{(3,3)}}                                                   ,   \underline{I_{i^{\prime}} \big( \underline{u} \big)}          \big\}   \big(  \underline{I_{i^{\prime\prime\prime}} \big( \underline{u} \big) }        \big) +                          \big( \underline{\mathcal{T}^{i^{\prime\prime}}_{(3,3)}}                     \big)               \big\{                                   \underline{I_{i^{\prime\prime\prime}} \big( \underline{u} \big) }         ,   \underline{I_{i^{\prime}} \big( \underline{u} \big)}          \big\}   \bigg]                             }                                   \bigg]      \text{. } 
            \end{align*} }

   \noindent For the first two terms of the superposition superposition above, $(*)$, there exists two strictly negative constants, $C_1$ and $C_2$, together implying that the first two terms can be approximated with,  
        
        {\small \begin{align*}
       (*)  \approx  - \underset{i,i^{\prime}, i^{\prime\prime} , i^{\prime\prime\prime}}{\sum}  \bigg[ {\tiny \frac{                     \underline{I_{i^{\prime\prime\prime}} \big( \underline{u} \big) }  \text{ }  \underline{I_{i^{\prime}} \big( \underline{u} \big)}       }{i^{\prime\prime}_{(3,3)} - i_{(3,3)} }   +      \frac{\big( \mathcal{T}^{i^{\prime\prime}}_{(3,3)}  \big)^{\prime}}{i^{\prime\prime\prime} - i^{\prime} } } \bigg]  \equiv  - \underset{i,i^{\prime}, i^{\prime\prime} , i^{\prime\prime\prime}}{\sum} 
 \bigg[ {\tiny \frac{                     \underline{
 I_{i^{\prime\prime\prime}} \big( \underline{u} \big) }  \text{ }  \underline{I_{i^{\prime}} \big( \underline{u} \big)}       }{i^{\prime\prime} - i }  +      \frac{\big( \mathcal{T}^{i^{\prime\prime}}_{(3,3)}  \big)^{\prime}}{i^{\prime\prime\prime} - i^{\prime} } }  \bigg]  \approx {\tiny \frac{C_1}{i^{\prime\prime}-i} + \frac{C_2}{i^{\prime\prime\prime} - i^{\prime}}   }       \text{, }  \end{align*}   }

        \noindent corresponding to the first, and last, bracket appearing in $(*)$, and another strictly negative constant,

          {\small  \begin{align*}
                C_{1,2,3}     \text{, }
            \end{align*} }

        \noindent corresponding to the remaining two brackets, from the observation,

            {\small \begin{align*}
                      {\tiny \big( \underline{\mathcal{T}^{i^{\prime\prime}}_{(3,3)}}                     \big)    }   \big\{                                   \underline{I_{i^{\prime\prime\prime}} {\tiny \big( \underline{u} \big) } }          , \underline{\mathcal{T}^i_{(3,3)}}         \big\} {\tiny \big(  \underline{I_{i^{\prime}} \big( \underline{u} \big)}            \big)  }  + {\tiny \big(   \underline{\mathcal{T}^i_{(3,3)}}  \big)      }                    \big\{           {\tiny     \underline{\mathcal{T}^{i^{\prime\prime}}_{(3,3)}}          }                                         ,   \underline{I_{i^{\prime}} {\tiny \big( \underline{u} \big)}}           \big\}  {\tiny \big(  \underline{I_{i^{\prime\prime\prime}} \big( \underline{u} \big) }        \big)   }  \approx \big[       {\tiny \big( \underline{\mathcal{T}^{i^{\prime\prime}}_{(3,3)}} +   \underline{\mathcal{T}^i_{(3,3)}}  \big)     }           \big]      \big[     \big\{   {\tiny \underline{I_{i^{\prime\prime\prime}}( \underline{u} )}}  , {\tiny \underline{\mathcal{T}^i_{(3,3)}}          }     \big\}  \\ + \big\{   {\tiny \underline{\mathcal{T}^{i^{\prime\prime}}_{(3,3)}}           }                                         ,   {\tiny \underline{I_{i^{\prime}} \big( \underline{u} \big)}      }    \big\} \big]  \big[ {\tiny \big( \underline{I_{i^{\prime}} \big( \underline{u} \big) } }         +  \underline{I_{i^{\prime\prime\prime}} {\tiny \big( \underline{u} \big) }   }     \big)       \big]  \approx   C_{1,2,3}  \text{. }
            \end{align*} }

\noindent Hence we obtain the desired sequence of constants appearing in the approximation,

{\small \[
\left\{\!\begin{array}{ll@{}>{{}}l} \boxed{(1)}:      \big\{  A {\tiny \big( \underline{u} \big)  }       , A {\tiny \big( \underline{u^{\prime}} \big) }   \big\} \approx  C_1 \equiv  \frac{                     (C_{1})_1   [ \underline{A ( \underline{u} )} ]  }{i^{\prime\prime} - i }  +      \frac{(C_{1})_2 [ \mathcal{T}_{(1,1))} ]}{i^{\prime\prime\prime} - i^{\prime}  }  + ( C_{1} )_3  
\text{ , } \\  \vdots  \\ \boxed{(81)}: \big\{ I {\tiny \big( \underline{u} \big) } , I {\tiny \big( \underline{u^{\prime}} \big) } \big\} \approx  C_{81} \equiv               \frac{                   ( C_{81} )_1 [ \underline{I ( \underline{u} )}] }{i^{\prime\prime} - i }  +      \frac{  ( C_{81} )_2 [ \mathcal{T}_{(3,3)} ] }{i^{\prime\prime\prime} - i^{\prime}  }   + ( C_{81} )_3   \text{ . }
\end{array}\right.
\]}

\noindent from which we conclude the argument. \boxed{}

\section{Conclusion}

\noindent In this paper we characterized integrability for the 20-vertex model under domain-wall boundary conditions. Through a suitably defined R-Matrix, the Universal R-Matrix, [5], we described how computations for product representations of L-operators can be used to asymptotically approximate the transfer and quantum monodromy matrices. The notion of weak integrability described for the 20-vertex model is related to that for the 6-vertex model through assumptions on the decay of several Poisson brackets of the block operators ${\tiny A \big( \underline{u} \big), \cdots, I \big( \underline{u} \big)}$.

\section{Appendix}

\subsection{Products of L-operators for the remaining column spans of the transfer matrix}

\subsubsection{Second column of the transfer matrix}

\noindent \textbf{Lemma} \textit{P2} (\textit{product representation of L-operators up to the fourth order}). The product of four L-operators is spanned by the union of three subspaces,

{\small \begin{align*}
\begin{bmatrix}
    D {\tiny \big( \underline{u}  \big)}  \\ E {\tiny \big( \underline{u} \big)}   \\ F {\tiny \big( \underline{u} \big)  }            
\end{bmatrix} \equiv \begin{bmatrix}
 4  \\ 5 \\ 6           
\end{bmatrix}  \text{, } 
\end{align*} }

\noindent for,

{\tiny \begin{align*}
    4 \equiv     \bigg[  \underline{\mathcal{T}^1_{(2,1)} }   + \underline{\mathcal{T}^2_{(2,1)}}    + \underline{\mathcal{T}^3_{(2,1)}}    + \underline{\mathcal{T}^4_{(2,1)}}               \bigg]  \bigg[  \underline{D_1 \big( \underline{u} \big) } + \underline{D_2 \big( \underline{u} \big) }  + \underline{D_3 \big( \underline{u} \big) }  + \underline{D_4 \big( \underline{u} \big) }  \bigg] \bigg[  \underset{i \in \textbf{N} :  |  i    | \leq  4 \lceil \frac{m-4}{3} \rceil }{\prod}  \underline{D_{1,i} \big( \underline{u}\big) }  \\    +    \underset{i \in \textbf{N} :  |  i    | \leq  4 \lceil \frac{m-4}{3} \rceil }{\prod}  \underline{D_{2,i} \big( \underline{u}\big) }   + \underset{i \in \textbf{N} :  |  i    | \leq  4 \lceil \frac{m-4}{3} \rceil }{\prod}  \underline{D_{3,i} \big( \underline{u}\big) } + \underset{i \in \textbf{N} :  |  i    | \leq  4 \lceil \frac{m-4}{3} \rceil }{\prod}  \underline{D_{4,i} \big( \underline{u}\big) }   \bigg]         \text{, } \end{align*}

  \begin{align*} 5 \equiv  \bigg[    \underline{\mathcal{T}^1_{(2,2)}} + \underline{\mathcal{T}^2_{(2,2)}}  + \underline{\mathcal{T}^3_{(2,2)}} + \underline{\mathcal{T}^4_{(2,2)}}  + \underline{\mathcal{T}^5_{(2,2)}} + \underline{\mathcal{T}^6_{(2,2)}}   \bigg] \bigg[ \underline{E_1 \big( u \big) }  + \underline{E_2 \big( \underline{u} \big) }  + \underline{E_3 \big( \underline{u} \big) }  + \underline{E_4 \big( \underline{u} \big) } +\underline{E_5 \big( \underline{u} \big) }  \\ +\underline{E_6 \big( \underline{u} \big) }   \bigg]   \bigg[  \underset{i \in \textbf{N} :  |  i    | \leq  4 \lceil \frac{m-4}{3} \rceil }{\prod}  \underline{E_{1,i} \big( \underline{u}\big) }  + \underset{i \in \textbf{N} :  |  i    | \leq  4 \lceil \frac{m-4}{3} \rceil }{\prod}  \underline{E_{2,i} \big( \underline{u}\big) }     + \underset{i \in \textbf{N} :  |  i    | \leq  4 \lceil \frac{m-4}{3} \rceil }{\prod}  \underline{E_{3,i} \big( \underline{u}\big) } \\  + \underset{i \in \textbf{N} :  |  i    | \leq  4 \lceil \frac{m-4}{3} \rceil }{\prod}  \underline{E_{4,i} \big( \underline{u}\big) }  + \underset{i \in \textbf{N} :  |  i    | \leq  4 \lceil \frac{m-4}{3} \rceil }{\prod}  \underline{E_{5,i} \big( \underline{u}\big) }      \bigg]        \text{, } \end{align*}

  \begin{align*}6 \equiv  \bigg[ \underline{\mathcal{T}^1_{(2,3)}} +  \underline{\mathcal{T}^2_{(2,3)}} + \underline{\mathcal{T}^3_{(2,3)}} \bigg] \bigg[  \underline{F_1 \big( \underline{u} \big) }       + \underline{F_2 \big( \underline{u} \big) } + \underline{F_3 \big( \underline{u} \big) }     \bigg]  \bigg[   \underset{i \in \textbf{N} :  |  i    | \leq  4 \lceil \frac{m-4}{3} \rceil }{\prod}  \underline{F_{1,i} \big( \underline{u}\big) }    \\   + \underset{i \in \textbf{N} :  |  i    | \leq  4 \lceil \frac{m-4}{3} \rceil }{\prod}  \underline{F_{2,i} \big( \underline{u}\big) }     +       \underset{i \in \textbf{N} :  |  i    | \leq  4 \lceil \frac{m-4}{3} \rceil }{\prod}  \underline{F_{3,i} \big( \underline{u}\big) }   \bigg]      \text{, }
\end{align*} }

 \noindent corresponding to the product,

{\small \[ \bigg\{ \begin{bmatrix}
  \textbf{1}^{1,\underline{j}} & \textbf{4}^{1,\underline{j}} & \textbf{7}^{1,\underline{j}}     \\   \textbf{2}^{1,\underline{j}} & \textbf{5}^{1,\underline{j}} & \textbf{8}^{1,\underline{j}}   \\ \textbf{3}^{1,\underline{j}} & \textbf{6}^{1,\underline{j}} & \textbf{9}^{1,\underline{j}} 
    \end{bmatrix}  \begin{bmatrix}
  \textbf{1}^{1,\underline{j}+1} & \textbf{4}^{1,\underline{j}+1} & \textbf{7}^{1,\underline{j}+1}     \\   \textbf{2}^{1,\underline{j}+1} & \textbf{5}^{1,\underline{j}+1} & \textbf{8}^{1,\underline{j}+1}   \\ \textbf{3}^{1,\underline{j}+1} & \textbf{6}^{1,\underline{j}+1} & \textbf{9}^{1,\underline{j}+1} 
    \end{bmatrix}  \begin{bmatrix}
  \textbf{1}^{1,\underline{j}+2} & \textbf{4}^{1,\underline{j}+2} & \textbf{7}^{1,\underline{j}+2}     \\   \textbf{2}^{1,\underline{j}+2} & \textbf{5}^{1,\underline{j}+2} & \textbf{8}^{1,\underline{j}+2}   \\ \textbf{3}^{1,\underline{j}+2} & \textbf{6}^{1,\underline{j}+2} & \textbf{9}^{1,\underline{j}+2} 
    \end{bmatrix}  \bigg\} \begin{bmatrix}
  \textbf{1}^{1,\underline{j}+3} & \textbf{4}^{1,\underline{j}+3} & \textbf{7}^{1,\underline{j}+3}     \\   \textbf{2}^{1,\underline{j}+3} & \textbf{5}^{1,\underline{j}+3} & \textbf{8}^{1,\underline{j}+3}   \\ \textbf{3}^{1,\underline{j}+3} & \textbf{6}^{1,\underline{j}+3} & \textbf{9}^{1,\underline{j}+3} 
    \end{bmatrix}   \times \cdots  \text{. } \] } 

\bigskip

\noindent \textit{Proof of Lemma P2}. Directly apply the argument for computing the product of L-operators provided in \textbf{Lemma} \textit{P1}, interchanging factors appearing in the factorization of the Universal R-Matrix where necessary, from which we conclude the argument. \boxed{}

\subsubsection{Third column of the transfer matrix}

\noindent \textbf{Lemma} \textit{P3} (\textit{product representation of L-operators up to the fourth order}). The product of four L-operators is spanned by the union of three subspaces,


{\small \begin{align*}
\begin{bmatrix}
   G {\tiny \big( \underline{u} \big)}  \\  H {\tiny \big( \underline{u} \big)}   \\ I {\tiny \big( \underline{u} \big)}        
\end{bmatrix} \equiv \begin{bmatrix}                          7 \\  8  \\  9 
\end{bmatrix}  , 
\end{align*} }

\noindent for,

{\tiny \begin{align*}
    7 \equiv      \bigg[ \underline{\mathcal{T}^1_{(3,1)}} +  \underline{\mathcal{T}^2_{(3,1)}}    + \underline{\mathcal{T}^3_{(3,1)}}  + \underline{\mathcal{T}^4_{(3,1)}}  + \underline{\mathcal{T}^5_{(3,1)}}   \bigg]   \bigg[  \underline{G_1 \big(\underline{u} \big)}     + \underline{G_2 \big( \underline{u} \big)}  + \underline{G_3 \big( \underline{u} \big)}  + \underline{G_4 \big( \underline{u} \big)} + \underline{G_5 \big( \underline{u} \big)} \bigg]  \bigg[ \underset{i \in \textbf{N} :  |  i    | \leq  4 \lceil \frac{m-4}{3} \rceil }{\prod}  \underline{G_{1,i} \big( \underline{u} \big) } \\ + \underset{i \in \textbf{N} :  |  i    | \leq  4 \lceil \frac{m-4}{3} \rceil }{\prod}  \underline{G_{2,i} \big( \underline{u} \big) }  + \underset{i \in \textbf{N} :  |  i    | \leq  4 \lceil \frac{m-4}{3} \rceil }{\prod}  \underline{G_{3,i} \big( \underline{u} \big) } + \underset{i \in \textbf{N} :  |  i    | \leq  4 \lceil \frac{m-4}{3} \rceil }{\prod}  \underline{G_{4,i} \big( \underline{u} \big) }  + \underset{i \in \textbf{N} :  |  i    | \leq  4 \lceil \frac{m-4}{3} \rceil }{\prod}  \underline{G_{5,i} \big( \underline{u} \big) } \bigg]   \text{, } \\ 8 \equiv      \bigg[       \underline{\mathcal{T}^1_{(3,2)}}   + \underline{\mathcal{T}^2_{(3,2)}}  + \underline{\mathcal{T}^3_{(3,2)}}  \bigg]   \bigg[   \underline{H_1 \big( \underline{u} \big) }   +  \underline{H_2 \big( \underline{u}\big) }   + \underline{H_3 \big( \underline{u} \big) }  \bigg]   \bigg[         \underset{i \in \textbf{N} :  |  i    | \leq  4 \lceil \frac{m-4}{3} \rceil }{\prod}  \underline{H_{1,i} \big( \underline{u}\big) } +  \underset{i \in \textbf{N} :  |  i    | \leq  4 \lceil \frac{m-4}{3} \rceil }{\prod}  \underline{H_{2,i} \big( \underline{u}\big) }  \end{align*}

 \begin{align*} +  \underset{i \in \textbf{N} :  |  i    | \leq  4 \lceil \frac{m-4}{3} \rceil }{\prod}  \underline{H_{3,i} \big( \underline{u}\big) } +  \underset{i \in \textbf{N} :  |  i    | \leq  4 \lceil \frac{m-4}{3} \rceil }{\prod}  \underline{H_{4,i} \big( \underline{u}\big) }          \bigg] \text{, } \\ 9 \equiv    \bigg[  \underline{\mathcal{T}^1_{(3,3)}} + \underline{\mathcal{T}^2_{(3,3)} }+ \underline{\mathcal{T}^3_{(3,3)}}  + \underline{\mathcal{T}^4_{(3,3)}        }     \bigg]  \bigg[   \underline{I_1 \big( \underline{u} \big)}  + \underline{I_2 \big( \underline{u} \big)}   + \underline{I_3 \big( \underline{u} \big)}    + \underline{I_4 \big( \underline{u} \big)}  
            \bigg]    
 \bigg[ \underset{i \in \textbf{N} :  |  i    | \leq  4 \lceil \frac{m-4}{3} \rceil }{\prod}  \underline{I_{1,i} \big( \underline{u}\big) }  +                      \underset{i \in \textbf{N} :  |  i    | \leq  4 \lceil \frac{m-4}{3} \rceil }{\prod}   \underline{I_{2,i} \big( \underline{u}\big) }  \\ +   \underset{i \in \textbf{N} :  |  i    | \leq  4 \lceil \frac{m-4}{3} \rceil }{\prod}  \underline{I_{3,i} \big( \underline{u} \big) } +                 \underset{i \in \textbf{N} :  |  i    | \leq  4 \lceil \frac{m-4}{3} \rceil }{\prod}  \underline{I_{4,i} \big( \underline{u} \big) }              \bigg]  \text{, }
\end{align*} }

 \noindent corresponding to the product,

{\small \[ \bigg\{ \begin{bmatrix}
  \textbf{1}^{1,\underline{j}} & \textbf{4}^{1,\underline{j}} & \textbf{7}^{1,\underline{j}}     \\   \textbf{2}^{1,\underline{j}} & \textbf{5}^{1,\underline{j}} & \textbf{8}^{1,\underline{j}}   \\ \textbf{3}^{1,\underline{j}} & \textbf{6}^{1,\underline{j}} & \textbf{9}^{1,\underline{j}} 
    \end{bmatrix}  \begin{bmatrix}
  \textbf{1}^{1,\underline{j}+1} & \textbf{4}^{1,\underline{j}+1} & \textbf{7}^{1,\underline{j}+1}     \\   \textbf{2}^{1,\underline{j}+1} & \textbf{5}^{1,\underline{j}+1} & \textbf{8}^{1,\underline{j}+1}   \\ \textbf{3}^{1,\underline{j}+1} & \textbf{6}^{1,\underline{j}+1} & \textbf{9}^{1,\underline{j}+1} 
    \end{bmatrix}  \begin{bmatrix}
  \textbf{1}^{1,\underline{j}+2} & \textbf{4}^{1,\underline{j}+2} & \textbf{7}^{1,\underline{j}+2}     \\   \textbf{2}^{1,\underline{j}+2} & \textbf{5}^{1,\underline{j}+2} & \textbf{8}^{1,\underline{j}+2}   \\ \textbf{3}^{1,\underline{j}+2} & \textbf{6}^{1,\underline{j}+2} & \textbf{9}^{1,\underline{j}+2} 
    \end{bmatrix}  \bigg\} \begin{bmatrix}
  \textbf{1}^{1,\underline{j}+3} & \textbf{4}^{1,\underline{j}+3} & \textbf{7}^{1,\underline{j}+3}     \\   \textbf{2}^{1,\underline{j}+3} & \textbf{5}^{1,\underline{j}+3} & \textbf{8}^{1,\underline{j}+3}   \\ \textbf{3}^{1,\underline{j}+3} & \textbf{6}^{1,\underline{j}+3} & \textbf{9}^{1,\underline{j}+3} 
    \end{bmatrix}  \times \cdots  \text{. } \] }

\noindent \textit{Proof of Lemma P3}. Directly apply the argument for computing the product of L-operators provided in \textbf{Lemma} \textit{P1}, interchanging factors appearing in the factorization of the Universal R-Matrix where necessary, from which we conclude the argument. \boxed{}

\section{Declarations}

\subsection{Availability of data and materials}

Not applicable

\subsection{Competing interests}

Not applicable

\subsection{Use of Generative AI}

Generative AI was not used to prepare or revised the submitted manuscript.

\section{References}

\noindent [1] Amico, L., Frahm, H., Osterloh, A., Ribeiro, G.A.P. Integrable spin-boson models descending from rational six-vetex models. \textit{Nucl. Phys. B.} \textbf{787}: 283-300 (2007). 
https://doi.org/10.1016/j.nuclphysb.2007.07.022.

\bigskip

\noindent [2] Alcaraz, F.C., Lazo, M.J. Exactly solvable interacting vertex models. \textit{J. Stat. Mech.} P08008 (2007). 
https://doi.org/

10.1088/1742-5468/2007/08/P08008.

\bigskip

\noindent [3] Batchelor, M.T., Baxter, R.J., O'Rourke, M.J., Yung, C.M. Exact solution and interfacial tension of the six-vertex model with anti-periodic boundary conditions. \textit{J. Phys. A: Math. Gen.} \textbf{28}: 2759 (1995). https://doi.org/10.1088/0305-4470/28/10/009.

\bigskip

\noindent [4] Baxter, R.J. Exactly Solved Models in Statistical Mechanics. Dover Publications, ISBN 9780486462714 (2018).

\bigskip

\noindent [5] Boos, H., et al. Universal R-matrix and functional relations. \textit{Rev. Math. Phys.} \textbf{26}: 143005 (2014). 
https://doi.org/10.1142/S0129055X14300052.

\bigskip

\noindent [6] Boos, H., Gohmann, F., Klumper, A., Nirov, K.S., Razumov, A.V.. Exercises with the universal R-matrix. J. Phys. A: Math. Theor. \textbf{43}: 415208 (2010).
https://doi.org/10.1088/1751-8113/43/41/415208.

\bigskip

\noindent [7] Colomo, F., Giulio, G.D., Pronko, A.G. Six-vertex model on a finite lattice: Integral representations for nonlocal correlation functions. \textit{Nuclear Physics B} \textbf{972}: 115535 (2021). 
https://doi.org/10.1016/j.nuclphysb.20

\noindent 21.115535.

\bigskip

\noindent [8] Colomo, F., Pronko, A.G. The Arctic Circle Revisited. \textit{Contemp. Math.} \textbf{458}: 361-376 (2008). https://www.ams.org/

books/conm/458/8947.

\bigskip

\noindent [9] Deguchi, T. Introduction to solvable lattice models in statistical and mathematical physics. \textit{CRC Press}: 9780429137891 (2003). https://doi.org/10.1201/9781420034

615.ch5.

\bigskip

\noindent  [10] Duminil-Copin, H., Karrila, A., Manolescu, I.,  Oulamara, M. Delocalization of the height function
of the six-vertex model. J. Eur. Math. Soc. (2024). https://ems.press/content/serial-article-files/48216.

\bigskip

\noindent [11] Duminil-Copin, H., Kozlowski, K.K., Krachun, D. et al. On the Six-Vertex Model’s Free Energy.
Commun. Math. Phys. 395, 1383–1430 (2022). https://doi.org/10.1007/s00220-022-04459-x.

\bigskip

\noindent [12] de Vega, H.J. Bethe Ansatz and Quantum Groups. \textit{LPTHE} \textbf{93/17} (1992). 
https://doi.org/10.48550/arXiv.hep-th/9308008.

\bigskip

\noindent [13] de Vega, H.J. Boundary K-matrices for the XYZ, XXZ and XXX spin chains. \textit{LPTHE-PAR} \textbf{93/29} (1993). 
https://doi.org/10.1088/0305-4470/27/18/021.

\bigskip

\noindent [14] de Vega, H.J., Ruiz, A.G. Boundary K-matrices for the six vertex and the $n \big( 2n-1 \big)$ $A_{n-1}$ vertex models. \textit{J. Phys. A: Math. Gen.} \textbf{26} (1993). 
https://doi.org/10.1088/0305-4470/26/12/007.

\bigskip

\noindent [15] Di Francesco, Philippe. Twenty Vertex model and domino
tilings of the Aztec triangle. the electronic journal of combinatorics \textbf{28}(4), (2021). https://doi.org/10.37236/10227.

\bigskip

\noindent [16] Faddeev, L.D., Takhtajan, L. A. Hamiltonian Methods in the Theory of Solitons. Springer Berlin, Heidelberg: Classics in Mathematics. ISBN 978-3-540-69843-2 (2007). https://doi.org/10.1007/978-3-540-69969-9.

\bigskip

\noindent [17] Frahm, H., Seel, A. The staggered six-vertex model: Conformal invariance and corrections to scaling. \textit{Nucl. Phys. B.} \textbf{879}: 382-406 (2014). 
https://doi.org/10.1016/j.nuclphysb.2013.12.015.

\bigskip

\noindent [18] Garbali, A., de Gier, J., Mead, W., Wheeler, M. Symmetric functions from the six-vertex model in half-space. \textit{arXiv: 2312.14348} (2023). 
https://doi.org/10.48550/arXiv.2312.14348.

\bigskip

\noindent [19] Gier, J.D., Korepin, V. Six-vertex model with domain wall boundary conditions: variable inhomogeneities. \textit{J. Phys. A: Math. Gen.} \textbf{34} (2001). 
https://doi.org/10.1088/0305-4470/34/39/312.

\bigskip

\noindent [20] Gohmann, F. Bethe ansatz. \textit{arXiv: 2309.02008} (2023). 
https://doi.org/10.48550/

arXiv.2309.02008.

\bigskip

\noindent [21] Gorbounov, V., Korff, C. Quantm integrability and generalised quantum Schubert calculus. \textit{Adv. Math.} \textbf{313}: 282-356 (2017). 
https://doi.org/10.1016/j.aim.2017.03.030.

\bigskip

\noindent [22] Gorbounov, V., Korff, C., Stroppel, C. Yang-Baxter Algebras as Convolution Algebras: The Grassmannian case. \textit{Russian Mathematical Surveys} \textbf{75} (2020). http://www.math.uni-bonn.de/ag/stroppel/Grass.pdf.

\bigskip

\noindent [23] Ikhlef, Y., Jacobsen, J., Saleur, H. A staggered six-vertex model with non-compact continuum limit. \textit{Nucl. Phys. B.} \textbf{789}(3): 483-524 (2008). https://doi.org/10.1016/j.nuclphysb.2007.07.004.

\bigskip

\noindent [24] Keating, D., Reshetikhin, N., Sridhar, A. Integrability of Limit Shapes of the Inhomogeneous Six Vertex Model. Commun. Math. Phys. 391, 1181–1207 (2022). https://doi.org/10.1007/s00220-022-04334-9

\bigskip

\noindent [25] Kitanine, N., et al. Thermodynamic limit of particle-hole-form factors in the massless XXZ Heisenberg chain. \textit{J. Stat. Mech. P05028} (2011). https://doi.org/10.1088/1742-5468/2011/05/P05028.

\bigskip

\noindent [26] Kitanine, N., et al. On correlation functions of integrable models associated with the six-vertex R-matrix. \textit{J. Stat. Mech.} P01022 (2007). 
https://doi.org/10.1088/1742-5468/2007/01/P01022.

\bigskip

\noindent [27] Kitanine, N., et al. On the spin-spin correlation functions of the XXZ spin-$\frac{1}{2}$ infinite chain. \textit{J. Phys. A: Math. Gen.} \textbf{38} (2005). 
https://doi.org/10.1088/0305-4470/38/34/001.

\bigskip

\noindent [28] Kozlowski, K.K. Riemann-Hilbert approach to the time-dependent generalized sine kernel. \textit{Adv. Theor. Math. Phys.} \textbf{15}: 1655-1743 (2011). https://doi.org/10.4310/ATMP.2011.v15.n6.a3.

\bigskip

\noindent [29] Korff, C., McCoy, B.M. Loop symmetry of integrable vertex models at roots of unity. \textit{Nucl. Phys. B.} \textbf{618}: 551-569 (2001). 
https://doi.org/10.1016/S0550-3213

\bigskip

\noindent [30]  Lieb, E. H. The Residual Entropy of Square Ice, Phys Rev. \textbf{162} 162-172 (1967). https://doi.org/10.1103/

\noindent PhysRev.162.162.

\bigskip

\noindent [31] Lamers, J. Introduction to quantum integrability. \textit{10th Modave Sumemr School in Mathematical Physics} (2015). https://doi.org/10.22323/1.232.0001.

\bigskip

\noindent [32] Motegi, K. Symmetric functions and wavefunctions of the XXZ-type six-vertex models and elliptic Felderhof models by Izergin-Korepin analysis. \textit{J. Math. Phys.} \textbf{59}: 053505 (2018). I:
https://doi.org/10.1063/

\noindent 1.4986534. 

\bigskip

\noindent [33] Naprienko, S. Free fermionic Schur functions. \textit{Adv. Math.} \textbf{436}: 109413 (2024). 
https://doi.org/10.1016/

\noindent j.aim.2023.109413. 

\bigskip

\noindent [34] Pauling, L. The Structure and Entropy of Ice and of Other Crystals with Some Randomness of Atomic Arrangement. J. Am. Chem. Soc. \textbf{57}, 2680 (1935). https://doi.org/10.1021/JA01315A102.

\bigskip

\noindent [35] Reshetikhin, N., Sridhar, A. Integrability of limits shapes of the six-vertex model. \textit{Comm. Math. Phys.} \textbf{356}: 535-565 (2017). https://doi.org/10.1007/s00220-017-2983-x.

\bigskip

\noindent [36]  Rigas, P. From logarithmic delocalization of the six-vertex height function under sloped boundary
conditions to weakened crossing probability estimates for the Ashkin-Teller, generalized random-cluster,
and $\big(q_{\sigma} , q_{\tau} \big)$-cubic models, \textit{arXiv:2211.14934}, Submitted (2024). 
https://doi.org/10.48550/arXiv.2211.14934.

\bigskip

\noindent [37] Rigas, P. Renormalization of crossing probabilities in the dilute Potts model. \textit{arXiv: 2211.10979}, Submitted (2026). https://doi.org/10.48550/arXiv.2111.10979.

\bigskip

\noindent [38] Rigas, P. The phase transition for the Gaussian free field is sharp. \textit{arXiv: 2307.12925}, Submitted (2023). 
https://doi.org/10.48550/arXiv.2307.12925.

\bigskip

\noindent [39] Rigas, P. Phase transition of the long range Ising model in lower dimensions, for $d < \alpha \leq d+1$, with a Peierls' argument. Accepted, to appear in International Journal of Modern Physics B.

\bigskip

\noindent [40] Rigas, P. Operator formalism for discreteley holomorphic parafermions of the two-color Ashkin-Teller, loop $\mathrm{O} \big( 1 \big)$, staggered eight-vertex, odd eight-vertex, and abelian sandpile models. \textit{arXiv: 2310.08212}, Submitted (2023). https://doi.org/10.48550/arXiv.2310.08212.

\bigskip

\noindent [41] Rigas, P. Poisson structure and Integrability of a Hamiltonian flow for the inhomogeneous six-vertex model. \textit{arXiv: 2310.15181}, Submitted (2023). https://doi.org/10.48550/arXiv.2310.15181.

\bigskip

\noindent [42] Rigas, P. Intertwining vectors, and Boltzmann weight matrices, of a Solid-on-Solid model from the 20-vertex model. Accepted, to appear in International Journal of Modern Physics B.

\bigskip

\noindent [43] Rigas, P. Scaling limit of the triangular prudent walk. \textit{arXiv: 2312.16236}, Submitted (2023). 
https://doi.org/

10.48550/arXiv.2312.16236.

\bigskip

\noindent [44] Rigas, P. Eigenvalue attraction in open quantum systems, biophysical systems, and Parity-time symmetric materials. \textit{arXiv: 2309.07943}, Submitted (2023). 
https://doi.org/10.48550/arXiv.2309.07943.

\bigskip

\noindent [45] Szabo, R.J., Tierz, M. Two-dimensional Yang-Mills theory, Painleve equations and the six-vertex model. \textit{J. Phys. A.: Math. Theor.} \textbf{45} (2012). 
https://doi.org/10.1088/1751-8113/45/8/085401.

\bigskip

\noindent [46] Tavares, T.S., Ribeiro, G.A.P. Finite temperature properties of an integrable zigzag ladder chain. \textit{Nucl. Phys. B.} \textbf{995} (2023). 
https://doi.org/10.1016/j.

nuclphysb.2023.116333.

\bigskip

\noindent [47] Zinn-Justin, P. Six-vertex model with domain wall boundary conditions and one-matrix model. \textit{Phys. Rev. E.} \textbf{62}, 3411 (2000). https://doi.org/10.1103/Phys

RevE.62.3411.

\bigskip

\noindent [48] Zinn-Justin, P. Six-vertex, Loop and Tiling Models: Integrability and Combinatorics. \textit{Lambert Academic Publishing}: 978-3-8383-2577-4 (2010).

\end{document}